\documentclass[
 reprint,
 amsmath,amssymb,
 aps,
]{revtex4-2}

\usepackage{graphicx}
\usepackage{dcolumn}
\usepackage{bm}

\usepackage[utf8]{inputenc}
\usepackage[english]{babel}
\usepackage[dvipsnames]{xcolor}
\usepackage{hyperref}

\usepackage{amsmath,amssymb}
\usepackage{mathtools}
\usepackage{mathrsfs}
\usepackage{physics}
\usepackage{dsfont}
\usepackage{svg}

\usepackage{soul}

\usepackage{fancyhdr}

\begin{document}


\title{A Finite-Time Quantum Otto Engine subject to Control Noise \\ and Enhancement Techniques}

\author{Theodore McKeever}
\author{Owen Diba}
\author{Ahsan Nazir}

\affiliation{Department of Physics and Astronomy, The University of Manchester, Manchester M13 9PL, UK.}
\date{\today}

\begin{abstract}
With the development of any quantum technology comes a need for precise control of quantum systems. Here, we evaluate the impact of control noise on a quantum Otto cycle. Whilst it is postulated that noiseless quantum engines can approach maximal Otto efficiency in finite times, the existence of white noise on the controls is shown to negatively affect average engine performance. Two methods of quantum enhancement, counterdiabatic driving and quantum lubrication, are implemented and found to improve the performance of the noisy cycle only in specified parameter regimes. To gain insight into performance fluctuations, projective energy measurements are used to construct a noise-averaged probability distribution without assuming full thermalisation or adiabaticity. From this, the variances in thermodynamic currents are observed to increase as average power and efficiency improve, and are also shown to be consistent with known bounds from thermodynamic uncertainty relations. Lastly, by comparing the average functioning of the unmonitored engine to a projectively-measured engine cycle, the role of coherence in work extraction for this quantum engine model is investigated.
\end{abstract}

\maketitle

\section{\label{sec:intro}Introduction}
The study of quantum engines is well established \cite{kosloff2017quantum,Alicki_1979,scovil1959three,feldmann2000performance,rezek2006irreversible}, providing useful theoretical frameworks to test the consistency of quantum and classical thermodynamics. For instance, the relationship between quantum and classical definitions of adiabaticity can also be viewed through the lens of quantum heat engines where the inclination of a quantum system to remain in its instantaneous eigenstate \cite{born1928va} directly influences the efficiency of the system operation. Quantum engines enable investigation into some unique features of thermodynamics in the quantum regime, such as overcoming classical performance trade-offs \cite{shiraishi2016universal} and the implementation of quantum enhancement techniques, both of which are addressed in this work. With recent physical realisations of quantum heat engines \cite{PhysRevLett.130.110402,PhysRevLett.122.110601,ExperimentalQHE,PhysRevA.108.042614}, there is also potential for functional applications, acting as sources of power for components within other quantum technologies. A cyclic heat engine operating in reverse functions as a refrigerator. Thus the findings presented here also apply indirectly to the performance of quantum refrigerators, which have a number of potential applications where reaching extremely cool and stable temperatures is required \cite{CANGEMI20241,Li_2021,PhysRevResearch.2.023120,PhysRevLett.132.210403}.

The presence of control noise is inevitable whenever classical technology is used to generate controls. Therefore, the inclusion of noisy parameters in quantum control is necessary in providing a complete analysis of any driven quantum system or realised quantum technology. Further, not just in practice but in principle, it has been posited that noise is fundamental to the interaction between quantum and classical objects due to the back-reaction of quantum systems on classical influences \cite{Layton2024healthiersemi}. In which case, accounting for control noise is vital for any theoretical model of a driven quantum system. The quantum engine analysed in this paper thus constitutes an open quantum system combining the influences of noisy driving protocols  and weak coupling to a thermal bath.

The utility of a quantum machine, like a classical machine, can be defined by its ability to complete a specific task. For instance, the charging of a battery by a (quantum) heat engine or the heat extraction from a cold bath by a (quantum) refrigerator. Such tasks often require the completion of several cycles. Therefore, the performance of quantum machines is commonly taken over many consecutive operations, where fluctuations are not usually considered and the cumulative action (some overall change in a thermodynamic observable) is measured \cite{kosloffCoherence}. This cycle is then repeated and divided by the number of cycle repetitions per experiment to calculate the average performance per cycle. In such cases, we calculate the average performance of a noisy quantum Otto engine using non-invasive energy expectation values. On the other hand, the full probability distribution is also valuable as it contains information on both average performance and fluctuations, and is particularly relevant, for example, under circumstances where quantum machines are needed only in short bursts with minimal consecutive operations. To construct the probability distribution for the quantum Otto engine, we probe the internal state of the working substance (a qubit) by subjecting it to projective energy measurements. Such measurements destroy coherence in the quantum system. Therefore, as well as accessing higher-order performance statistics for individual cycles, we are able to compute the role of coherence in the quantum engine by comparing results from projective measurements versus non-invasive expectation values.

The paper is organised as follows. In Section~\ref{sec:model} we introduce the engine model under review, including details of the cycle strokes and the existence of a limit cycle. We then implement methods of performance enhancement within the quantum regime in Section~\ref{sec:enhance}, and assess the average performance of the engine with and without such techniques in Section~\ref{sec:avgPerf} and over different stroke durations in Section~\ref{sec:durations}. Next, we calculate the full probability distribution of an equivalent engine using invasive energy measurements within the cycle. From this, we deduce the expected performance and its variance in Section~\ref{sec:fluctuations}, before summarising and drawing our conclusions in Section~\ref{sec:conc}.



\section{\label{sec:model}Engine Model}
\subsection{\label{subsec:otto}The Quantum Otto Cycle}

\begin{figure}
    \centering
    \includegraphics[width=0.95\linewidth]{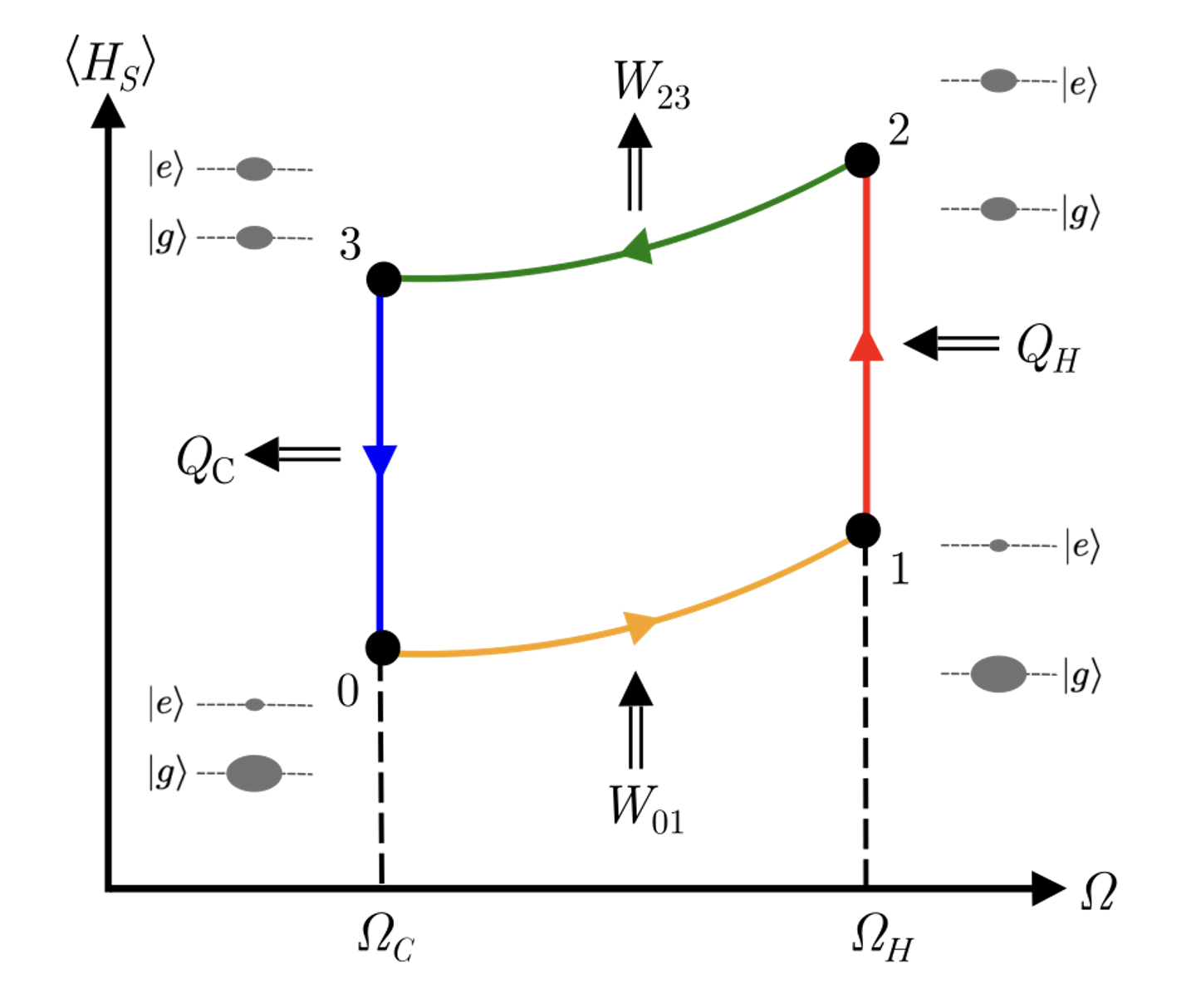}
 
\caption{\label{fig:ottoCycle} In the 4-stroke Otto engine, the system exchanges heat $Q_i$ ($i=H,C$) with a bath at inverse temperature $\beta_i$, along the isochores (Red, Blue). Work $W_{01}$ is performed on the system during isentropic compression (Orange) and work $W_{23}$ is extracted from the engine during isentropic expansion (Green). $\Omega$ represents the energy splitting of the system and $\langle H_S \rangle $ represents the expectation value of the internal energy of the system. Approximate energy level diagrams are given for each cycle vertex, showing the change in populations and energy level splitting in the energy eigenbasis $\{ \ket{e} ,\ket{g} \}$.}
\end{figure}

In parallel to their classical counterparts, there are several types of quantum engines operating either continuously \cite{kosloff2014quantum} or in a reciprocating cycle. The Otto engine is a paradigmatic example of the latter, consisting of four strokes: two isentropes and two isochores, performed cyclically. Each stroke $a \rightarrow b$ has an allocated time $\tau_{ab}$ which for the remainder of this work are taken to be equal to each other. Explicitly, the stages  (illustrated in Figure~\ref{fig:ottoCycle}) are as follows:
\begin{description}
     \item[$0 \rightarrow 1$] \textbf{Compression.} On this isentrope, work is done on the system ($W_{01}$), altering the system Hamiltonian from $H_C \rightarrow H_H$, increasing the difference between energy eigenvalues, $\Omega_C \rightarrow \Omega_H$ as in Equations~\ref{eqn:ham1} and \ref{eqn:ham2} below.
     \item[$1 \rightarrow 2$] \textbf{Heating.} On this isochore, the system is coupled to the hot bath at inverse temperature $\beta_H$, absorbing heat $Q_H$ whilst the Hamiltonian is kept constant, $H_H$.
     \item[$2 \rightarrow 3$] \textbf{Expansion.} On this isentrope, work is done by the system ($-W_{23}$), altering the system Hamiltonian from $H_H \rightarrow H_C$, which decreases the difference between energy eigenvalues, $\Omega_H \rightarrow \Omega_C$.
     \item[$3 \rightarrow 0$] \textbf{Cooling.} On this isochore, the system is coupled to the cold bath at inverse temperature $\beta_C$, releasing heat $Q_C$ whilst the Hamiltonian is kept constant, $H_C$.
\end{description}

\subsection{\label{subsec:opendynamics}Open Quantum Dynamics}

The working substance of the quantum Otto engine is taken to be a two-level system with intrinsic (and constant) tunnelling, $\Omega_x\sigma_x$, and with an applied external control field providing a time-dependent energy level splitting, $\Omega_z(t)\sigma_z$. This could represent a variety of physical systems, for instance, trapped ions with energy level splitting driven by laser pulses \cite{PhysRevA.63.012306,Roos2014}, superconducting circuits \cite{PhysRevB.94.184503,chiorescu2003coherent,PhysRevA.90.022307,PRXQuantum.2.040204}, or quantum dots controlled by electromagnetic fields \cite{Ulhaq13,PhysRevB.108.195307,abo2020long}. This setup is described by the system Hamiltonian
\begin{equation}
    \label{eqn:ham1}
    H_S(t) = \frac{1}{2}\left( \Omega(t)\mathds{1} + \Omega_x \sigma_x + \Omega_z(t)\sigma_z \right)
\end{equation}
where the identity term shifts the eigenvalues to $\Omega(t)$ and $0$, with $\Omega(t)=\sqrt{\Omega_x^2+\Omega_z(t)^2}$. $H_S(t)$ can be represented in its energy eigenbasis (we denote any operator in this basis by a prime) by rotating with $R(t)=\text{e}^{-i \theta(t) \sigma_y/2}$, where $\theta (t) = \arctan\left( \Omega_x/\Omega_z(t)\right)$:
\begin{equation}
    \label{eqn:ham2}
    H_S'(t) = R^\dagger (t) H_S(t) R(t) = \begin{pmatrix}
        \Omega(t) & 0 \\
        0 & 0 
        \end{pmatrix}.
\end{equation}
The system is driven between $H_C \leftrightarrow H_H$ as $\Omega_z(t)$ is varied. A polynomial ansatz is used for the sweeping of $\Omega_z(t)$ such that it varies smoothly across the isentropes between $\Omega_{z,C} \leftrightarrow \Omega_{z,H}$ with a zero time-derivative at the beginning and end of each stroke. In terms of dimensionless time, $s_{ab} = t_{ab}/\tau_{ab}$, over an isentrope $a \rightarrow b$, the form of the external driving is given by $\Omega_z(t)= (3s_{ab}^2-2s_{ab}^3)(\Omega_{z}^{b}-\Omega_{z}^a)+\Omega_{z}^a$ where $\Omega_{z}^a$ ($\Omega_{z}^{b}$) is the chosen value of $\Omega_z$ at the beginning (end) of the stroke \cite{PhysRevE.99.022110}.

Along the isochores, the hot and cold thermal reservoirs can be modelled as large, memoryless bosonic baths weakly interacting with a static system:
    \begin{align}\label{eqn:HamStatTot}
    H_T & = H_S + H_B + H_I \nonumber\\
    & = H_S + \sum_k\omega_k b_k^\dagger b_k + \sigma_z \otimes \sum_k g_k(b_k^\dagger +b_k).
    \end{align}
Here $H_B$ is given as a sum of quantum harmonic oscillators \cite{hofer2017markovian}, where $b_k^\dagger$ and $b_k$ are the raising and lowering operators of the bath modes with frequencies  $\omega_k$, which couple to the system with strengths $g_k$.

Within the Born-Markov approximations, the von Neumann equation $\Dot{\rho}(t) = -i[H(t),\rho(t)]$ that defines the unitary evolution of a closed quantum system is then adapted to include a bath dissipator, $\mathcal{D}_B$, which describes the interaction of the system with the hot and cold baths and introduces non-unitary behaviour into the system dynamics. The resulting well-established GKLS equation has the form \cite{breuertheory}
\begin{equation}
    \label{eqn:GKLS}
    \Dot{\rho}(t) = -i[H(t),\rho(t)] + \mathcal{D}_B (\rho(t))
\end{equation} 
with
\begin{equation}
    \label{eqn:bathdissipator}
    \mathcal{D}_B(\rho(t)) = \sum_{n = +,-,0} \gamma_n(P_n\rho(t) P_n^{\dagger} - 1/2\{P_n^{\dagger}P_n, \rho(t)\}),  
\end{equation} 
where $P_{+,-,0}$ are jump operators ($P_+=\ketbra{e}{g}$, $P_-=\ketbra{g}{e}$, and $P_0=\ketbra{e}{e}-\ketbra{g}{g}$) and $\gamma_n$ are the corresponding Lindblad rates. The rates scale with the bath spectral density, $J(\omega)$, and the occupation number $N(\Omega) = 1/(e^{\Omega \beta}-1)$ where $\beta$ is the inverse temperature of the bath: $\gamma_+ =  J(\omega) N(\Omega) 2\pi\cos^2 \theta$, $\gamma_- = J(\omega)\left(1+ N(\Omega)\right) 2\pi \cos^2 \theta$ and $\gamma_0 = 2\pi (\alpha/\beta) \sin^2 \theta $ ($\alpha$ is the system-bath coupling strength). The spectral density describes the distribution of frequencies at which the system interacts with its environment and is chosen here to be Ohmic, $J(\omega) = \alpha \omega e^{-\omega/\omega_C}$ ($\omega_C$ is the cut-off frequency).

To summarise, the system is weakly-coupled to a heat bath on the isochores, but not driven. Conversely, on the isentropes, the system is treated in isolation from the thermal environments but the energy splitting is driven using $\Omega_z(t)$.

\subsection{\label{subsec:noisydynamics} Noisy Quantum Dynamics}
Control noise during the cycle is modelled as a Gaussian White Noise (GWN) process, $\xi_i(t)$, with the following properties \cite{RevModPhys.82.1155}: (i) noise values obey a Gaussian distribution centred on zero, $\mathds{E}[\xi_i(t)]=0$; (ii) the noise is delta-correlated such that there are no correlations between different noise sources nor with the same source at different times, $\mathds{E}[\xi_i(t)\xi_j(s)]=\lambda \mathcal{X}_{i j}(t-s)$ where $\lambda$ characterises the noise strength and $\mathcal{X}_{i j}(t-s)= \delta_{ij}\delta(t-s) $ defines the two-time correlations; (iii) there is no initial correlation, $\mathds{E}[\xi_i(0)\rho(0)]=0$. Here, $\mathds{E}[...]$ represents an average over noise-realisations and $\delta$ the Kronecker delta.

Noise is implemented such that it scales with the amplitude of any associated control fields:
\begin{equation}
    \label{eqn:noisyHam}
    H_{\xi}(t)=\sum_j\frac{1}{2}\xi_j(t)\Omega_j(t)\sigma_j .
\end{equation}
The noisy system Hamiltonian is the sum of $H_S(t)$ from Equation~(\ref{eqn:ham1}) and $H_{\xi}(t)$, $H_{S,\xi}(t)=H_S(t) + H_{\xi}(t)$. From this and the properties of GWN, (i)-(iii), the noise-averaged dynamics governed by the dissipator $\mathcal{D}_\xi(\mathds{E}[\rho(t)])$ can be derived (see Appendix \ref{sec:appendixExp}) utilising Novikov's theorem \cite{novikov1965functionals} and following the approach in \cite{noiseFactorisation}, arriving at:
\begin{equation}
    \label{eqn:noisyDissipator}
    \mathcal{D}_\xi(\mathds{E}[\rho(t)]) = - \sum_j \frac{\Omega_j(t)^2}{4} \lambda[\sigma_j,[\sigma_j,\mathds{E}[\rho(t)]]].
\end{equation}

The general state evolution equation for the cycle includes both unitary evolution, and additive non-unitary contributions from noisy controls and weak coupling to a heat bath (hot or cold),
    \begin{align}\label{eqn:evolution1}
    \mathds{E}\left[ \Dot{\rho}(t)\right] = & -i \left[ H_S(t),\mathds{E}[\rho(t)]\right] + \mathcal{D}_{B}(\mathds{E}[\rho(t)])\nonumber\\
    & +\mathcal{D}_{\xi}(\mathds{E}[\rho(t)]).
    \end{align}
More specifically, $\mathcal{D}_{B}(\mathds{E}[\rho(t)])$ is not present on isentropes and $H_S(t)$ is time-independent on isochores. Thus, whilst $\mathcal{D}_{\xi}(\mathds{E}[\rho(t)])$ is present on all strokes, it is only time-dependent on the isentropes, where $\Omega_j(t)$ varies with time. 

In order to gather results for the performance of the engine, it is necessary to employ a definition of heat and work for noisy strokes. Here, we consider all energy changes across thermal strokes (isochores) as heat exchanges and all energy changes on the driving strokes (isentropes) as work exchanges, as would be recorded by an observer with access only to projective measurements on the working system. A full discussion of our reasoning on this matter, including alternative definitions with corresponding calculations, can be found in Appendix~\ref{sec:appendixDef}. Note that the terms \textit{isochore} and \textit{isentrope} will continue to be used to refer to work and heat strokes, despite the presence of noise compromising their original meanings.



\subsection{\label{subsec:limitcycle}The Limit Cycle}
The focus of this work is on finite-time cycles. Therefore, it would not be appropriate to assume full thermalisation on the isochores, nor adiabatic dynamics on the isentropes. Having been prepared in any input state, after $N$ repetitions of the cycle, the system reaches its limit cycle where the state at a given vertex, $m$, of consecutive cycles is the same, $\rho_N^m=\rho^m_{N-1}$. The existence of a limit cycle state for a given set-up can be proved by defining the total cycle propagator, $\mathcal{V}_{cyc}$, and finding the state, $\rho_\text{lim}$, on which there is no net effect (the eigenvector of $\mathcal{V}_{cyc}$ with eigenvalue of 1), in vectorised form $\mathcal{V}_{cyc} \ket{\rho_\text{lim}}\rangle =  \ket{\rho_\text{lim}}\rangle$ \cite{limitCyclesIsinga}. All engines considered here have a unique (noise-averaged) limit cycle state, which has been confirmed by examination of the transient dynamics approaching the limit cycle from initially thermal and collapsed states. All results quoted are of engines operating in their finite-time limit cycles.


\section{\label{sec:enhance}Enhancement Techniques }
In thermodynamics, adiabaticity refers to the minimisation of dissipation and, hence, entropy production. Adiabatic processes have maximal efficiency; for engines, this is equal to the Carnot efficiency, $\eta_C = 1- T_C/T_H$ \cite{kosloff2017quantum}. Ordinarily this optimal efficiency can only be achieved in the infinite time-limit, where power output is zero. Thus, there is a trade-off between power and efficiency. Under the assumptions of perfect thermalisation on isochores and noiseless adiabatic driving on isentropes, the maximum work output and heat of the Otto engine can be found by taking the energy expectation value of the system at each vertex over one cycle, yielding
    \begin{align}\label{eqn:MaxWork}
    & W_{max}=\frac{\Omega_H-\Omega_C}{2}\left( \tanh{\frac{\beta_C \Omega_C}{2}}-\tanh{\frac{\beta_H \Omega_H}{2}} \right) , \\
    & Q_{H,max} = \frac{\Omega_H}{2} \left( \tanh{\frac{\beta_C \Omega_C}{2}}-\tanh{\frac{\beta_H \Omega_H}{2}} \right).
    \end{align}
Thus, the optimal Otto efficiency in quantum engines can be calculated as $\eta_{O} = 1- \Omega_C/\Omega_H$. This is equivalent to $\eta_C$ in the instance where $\Omega_C/\Omega_H=T_C/T_H$, in which case there is zero work output. Thus, a trade-off between efficiency and power exists for noiseless quantum engines with respect to the driving parameter $\Omega(t)$. Further, as in the classical case, a similar trade-off may be expected with respect to the choice of $\tau$, the cycle operation time: as $\tau \rightarrow \infty$, power output approaches zero and efficiency becomes maximal. 

When the system is driven over a finite time and the Hamiltonian is not two-time-commuting, off-diagonal non-adiabatic correction terms appear in the 
eigenbasis of the system Hamiltonian evolution: 
     \begin{align}\label{eqn:friction}
     \Dot{\rho}'(t)& =\frac{d}{dt}\left( R^\dagger (t) \rho (t) R(t) \right) \nonumber\\
     & = -i \left[H_S'(t), \rho' (t)\right] + i\left[\frac{\Dot{\theta}(t) \sigma_y}{2}, \rho'(t)\right].
    \end{align}
The dynamics induced by such terms generate coherences in the system energy eigenbasis and have an associated energetic cost, coined \textit{quantum friction} \cite{PhysRevE.65.055102}.

Methods of quantum enhancement can, in theory, be implemented to counteract quantum friction in the system and thus to recover adiabatic behaviour in finite operation times. One approach is to apply a \textit{shortcut to adiabaticity} (STA) \cite{PhysRevX.4.021013}. One widely applicable STA is counterdiabatic driving which involves applying an auxiliary field to the system that directly cancels the non-adiabatic correction terms \cite{guery2019shortcuts,berry2009transitionless}. Here, the Hamiltonian for counterdiabatic driving on the engine isentropes is $H_{STA}(t)=\frac{1}{2}\Omega_y(t)\sigma_y$, where $\Omega_y(t)=\Dot{\theta}(t)=-\frac{\Omega_x\Dot{\Omega}_z(t)}{\Omega_x^2+\Omega_z(t)^2}$. When noise is considered, the externally-driven STA field will also be subject to amplitude noise in the same fashion as other controls (Equation~\ref{eqn:noisyHam}).

Another approach to avoid quantum friction effects is \textit{quantum lubrication} (QL) \cite{QLubrication}, where a pure-dephasing (noisy) field is applied to the system. A noise source is pure-dephasing with respect to an observable $X$ if its dissipator (derived from Equation~\ref{eqn:noisyDissipator}) acts only and negatively on the coherences (off-diagonals) of the system density matrix in the eigenbasis of $X$. This is achieved for the energy observable $H_S(t)$ when the noisy lubricating field is proportional to $\sigma_z'(t)$ in the $H_S(t)$ eigenbasis ($ \sigma_z'(t) = (\Omega_z(t) \sigma_z + \Omega_x \sigma_x)/ \Omega(t)$ in the unrotated basis). An applied field of $H_{QL}'(t)=\xi_{QL}(t)\sigma_z'(t)$, where $\xi_{QL}(t)$ is a GWN source with controllable noise strength $\lambda_{QL}(t)$, also ensures that the original Hamiltonian (Equation~\ref{eqn:ham1}) is returned after noise-averaging. Here, we take $\lambda_{QL}$ to be constant. The pure-dephasing dissipator then acts as 
     \label{eqn:QL}
     \begin{align}
     \mathcal{D}_{QL}(\mathds{E}[\rho'(t)]) & = - \lambda_{QL}[\sigma_z'(t),[\sigma_z'(t),\mathds{E}[\rho'(t)]]]\nonumber\\
     & = -4\lambda_{QL} \begin{pmatrix}
        0 & \rho'_{01}(t) \\
        \rho'_{10}(t) & 0 
        \end{pmatrix},
    \end{align}
where $\rho'_{ij}(t)$ is the entry on the $i^\text{th}$ row and $j^\text{th}$ column of the system state, $\rho(t)$, expressed in the (noise-averaged) energy eigenbasis.

Thus, Equation~\ref{eqn:evolution1} can be adapted for enhanced isentropes (which are isolated from all heat reservoirs) either by QL,
    \begin{align}\label{eqn:evolutionQL}
    \mathds{E}\left[ \Dot{\rho}(t)\right] = & -i \left[ H_S(t),\mathds{E}[\rho(t)]\right] +\mathcal{D}_{\xi}(\mathds{E}[\rho(t)])  \nonumber\\
    & +\mathcal{D}_{QL}(\mathds{E}[\rho(t)]),
    \end{align}
or by an STA, 
    \begin{align}\label{eqn:evolutionSTA}
    \mathds{E}\left[ \Dot{\rho}(t)\right] = & -i \left[ H_S(t) + H_{STA}(t),\mathds{E}[\rho(t)]\right] \nonumber\\
    & +\mathcal{D}_{\xi}(\mathds{E}[\rho(t)]),
    \end{align}
where in the latter case $\mathcal{D}_{\xi}(\mathds{E}[\rho(t)])$ is adapted to include an extra noise term proportional to $H_{STA}(t)$. 
\section{\label{sec:avgPerf} Average Performance}

\begin{figure}
    \centering
    \includegraphics[width=0.47\textwidth]{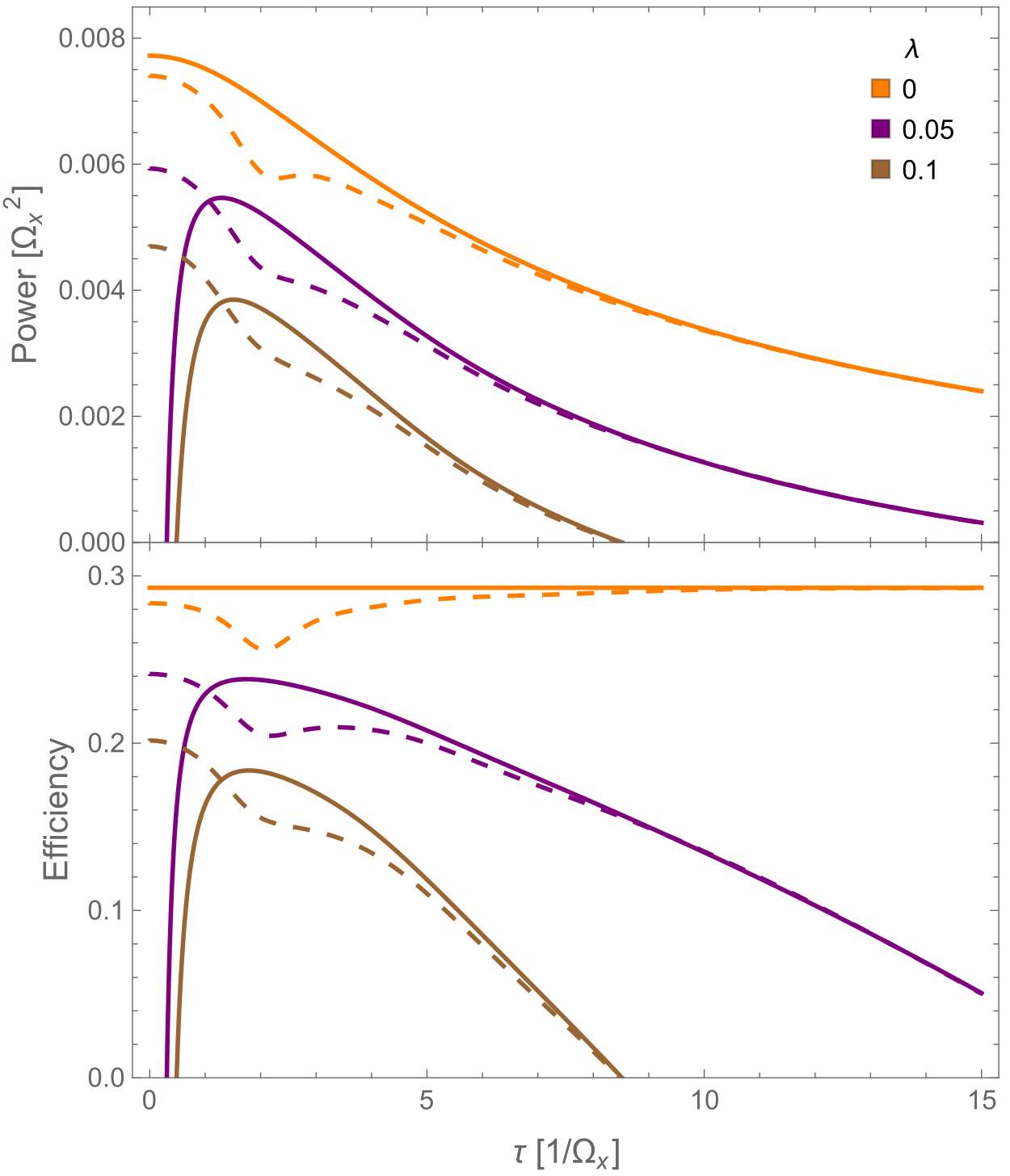}
\caption{\label{fig:AvgPerformance} The power (top) and efficiency (bottom) of the engine in the presence of three different levels of noise, $\lambda$, across different cycle times, $\tau$, given in units of intrinsic system energy scales, $\Omega_x$. In all plots, the solid lines represent an engine with an STA and the dashed lines without. The environment parameters are: $\alpha=0.01$, $\Omega_C=100$, $T_C=T_H/5=10\Omega_x$.}
\end{figure}

We are now in a position to assess the average performance of the Otto engine model described in the previous sections. In this section, we will compare an unenhanced, nonadiabatic (NA) engine with its corresponding STA and QL counterparts, where results are generated by numerically solving the dynamics along successive strokes of equal duration. We then generalise these results to other proportions of isochoric and isentropic stroke times in the next section. 


Looking at the NA engine operation (Figure~\ref{fig:AvgPerformance}, dashed lines), control noise is seen to routinely worsen engine performance. In the noiseless case (orange), power is maximised at short cycle times, $\tau$, whilst efficiency is maximised as $\tau \rightarrow \infty$, thus exhibiting the classical trade-off between power and efficiency. With noise, the NA engine has maximal power and efficiency under a pseudo-bang-bang control strategy (near instantaneous driving).

When an STA is applied (solid lines), the efficiency is maximised at the Otto efficiency in the absence of control noise, where $\eta(\tau)=\eta_O ~\forall \tau$. When control noise is present, power and efficiency peak at similar, but not exactly equivalent, finite cycle times. STAs are found to improve engine performance with diminishing returns at larger cycle times, demonstrating the legitimacy of STAs in recapturing adiabatic dynamics. Notably, with noisy controls the power and efficiency of engines with STAs are degraded at small $\tau$ below a critical operation time, $\tau_\text{STA}$. More generally, counterdiabatic driving is expected to be ineffective for fast driving with amplitude-noise. This is because $H_{STA}(t)$ will act to counteract non-adiabatic correction terms (proportional to $\Dot{\Omega}_z(t)$); meaning the faster the controls are swept, the greater the magnitude of $H_{STA}(t)$. Therefore, since amplitude noise scales with the magnitude of the control field, the STA field is more affected by noise when the system is driven quickly. Looking at an individual trajectory this means that $H_{STA}(t)$ is highly inexact and fails to cancel the non-adiabatic correction terms. When looking at the noise-averaged dynamics, this corresponds to a large $\mathcal{D}_\xi(\mathds{E}[\rho(t)])$ associated with the STA field. Since the intended applications for STAs are usually at shorter timescales, the existence of $\tau_\text{STA}$ is noteworthy. Nevertheless, counterdiabatic driving is predicted to improve efficiency and power at operation times larger than $\tau_\text{STA}$. The time domain where STAs are effective is broadened as $\tau_\text{STA}$ is reduced, which can be achieved by lowering noise levels, $\lambda$, or increasing $\Omega(t)/\Omega_x$. For experimental realisations with energies of the order of 1~GHz, effective timescales for STA use are $\tau_\text{STA} \approx 1/\Omega_x = 1 ~\text{ns}$.

\begin{figure}
    \centering
    \includegraphics[width=0.47\textwidth]{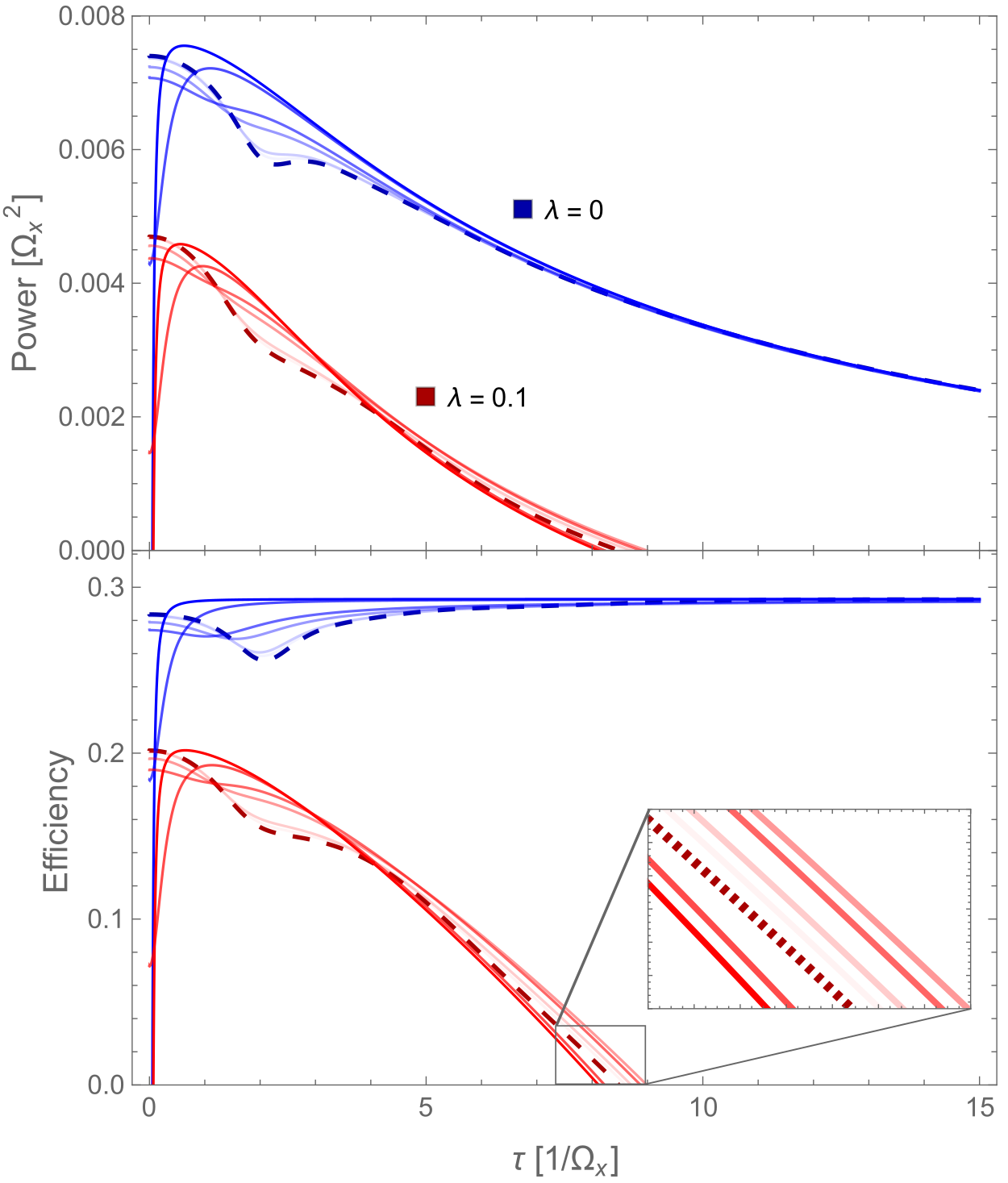}
    
\caption{\label{fig:QL} The power (top) and efficiency (bottom) of the quantum Otto cycle with different strengths of lubricating noise, $\lambda_{QL}$ (solid), with (red) and without (blue) control noise, $\lambda$. This is shown in comparison to NA performance (dashed) across different cycle times, $\tau$, given in units of intrinsic system energy scales, $\Omega_x$. $\lambda_{QL}=\{0.05, 0.1, 0.5, 1, 10, 100\}$ increasing with degree of colour saturation. The environment parameters are: $\alpha=0.01$, $\Omega_C=100$, $T_C=T_H/5=10\Omega_x$.}
\end{figure}

Turning to Figure~\ref{fig:QL}, in the noiseless case QL is successful in recovering adiabatic dynamics at finite time when the dephasing noise is strong (darker solid blue lines). With weak dephasing noise (lighter blue), performance improves but remains sub-optimal, except at very short times when retaining coherence is advantageous \cite{kosloffCoherence} (see also the next section). When control noise is present, unlike STA engines, those with QL can retain positive performance even at very small $\tau$ (weak QL), and can also have a larger maximum power production than STA engines (strong QL). However, at larger $\tau$, the QL engine behaves non-trivially. Depending on $\lambda_{QL}$, QL can worsen engine performance or improve it beyond adiabatic predictions. This is likely due to the fact that, in the presence of noise, the instantaneous eigenbasis of the Hamiltonian is fluctuating. Therefore, the dephasing noise, which is applied diagonally in the noise-averaged eigenbasis, no longer commutes with the system Hamiltonian. Equivalently, in the noise-averaged formalism, $\mathcal{D}_{QL}(\mathds{E}[\rho'(t)])$ interferes with the action of $\mathcal{D}_{\xi}(\mathds{E}[\rho'(t)])$ by destroying its off-diagonal contribution to $\Dot{\rho}'(t)$, as well as non-adiabatic Hamiltonian contributions. Thus, the dephasing field can give rise to non-intuitive energetic contributions. For instance, using parameters from Figure~\ref{fig:QL}, at large $\tau$ there is an optimal value for $\lambda_{QL}$ ($\approx 4$) up to which performance improves above that of the NA engine and beyond which performance reduces.

Despite its detrimental behaviour at both very small and large $\tau$, in the remaining analysis and comparison between approaches we continue with strong QL ($\lambda_{QL}=100$) due its advantages for other $\tau$ values.


\section{\label{sec:durations} Stroke Durations}

\begin{figure*}
    \centering
    \begin{minipage}{0.485\textwidth}
        \centering
        \includegraphics[width=\linewidth]{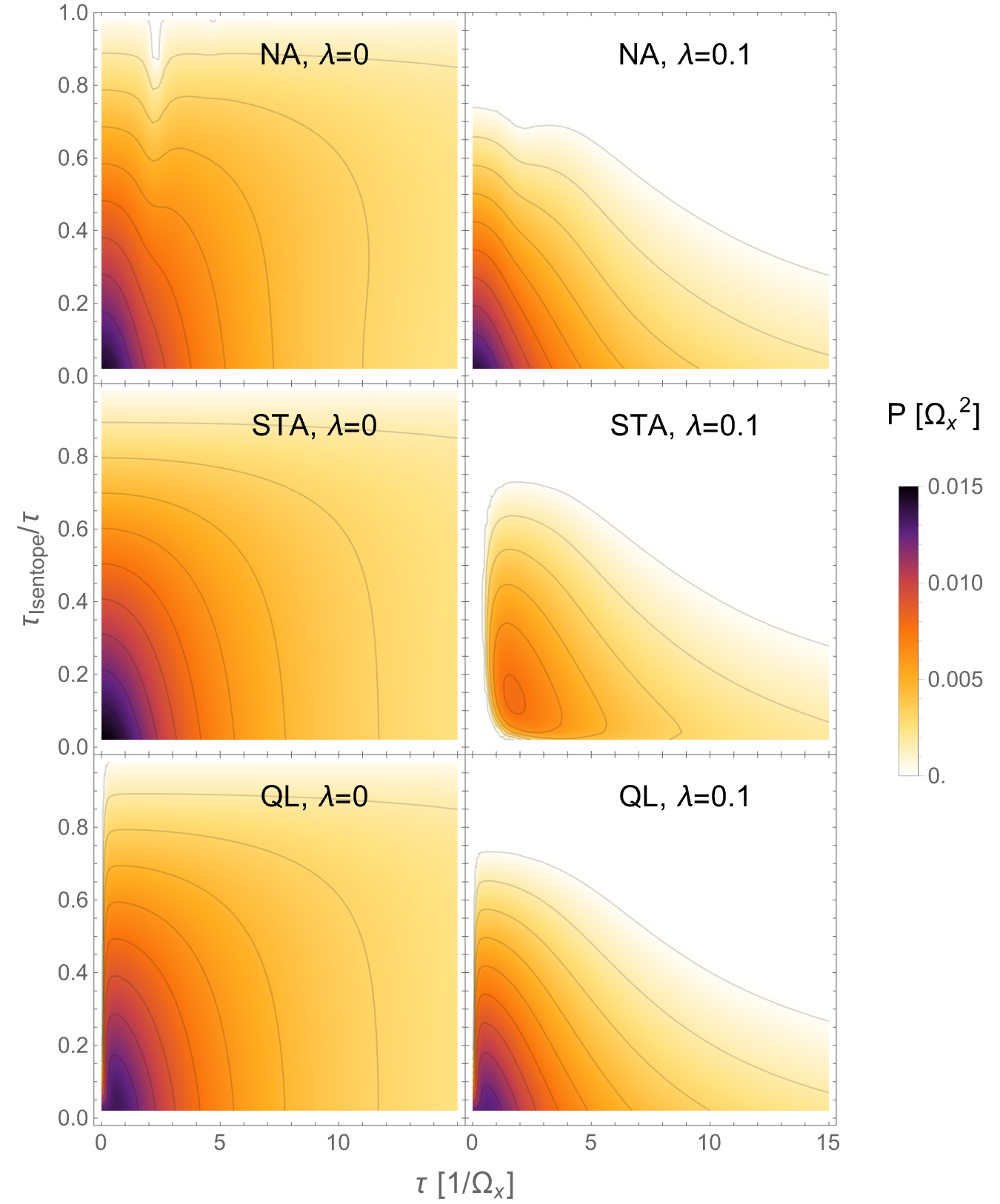}
    \end{minipage}
    \hfill
    \begin{minipage}{0.50\textwidth}
        \centering
        \includegraphics[width=\linewidth]{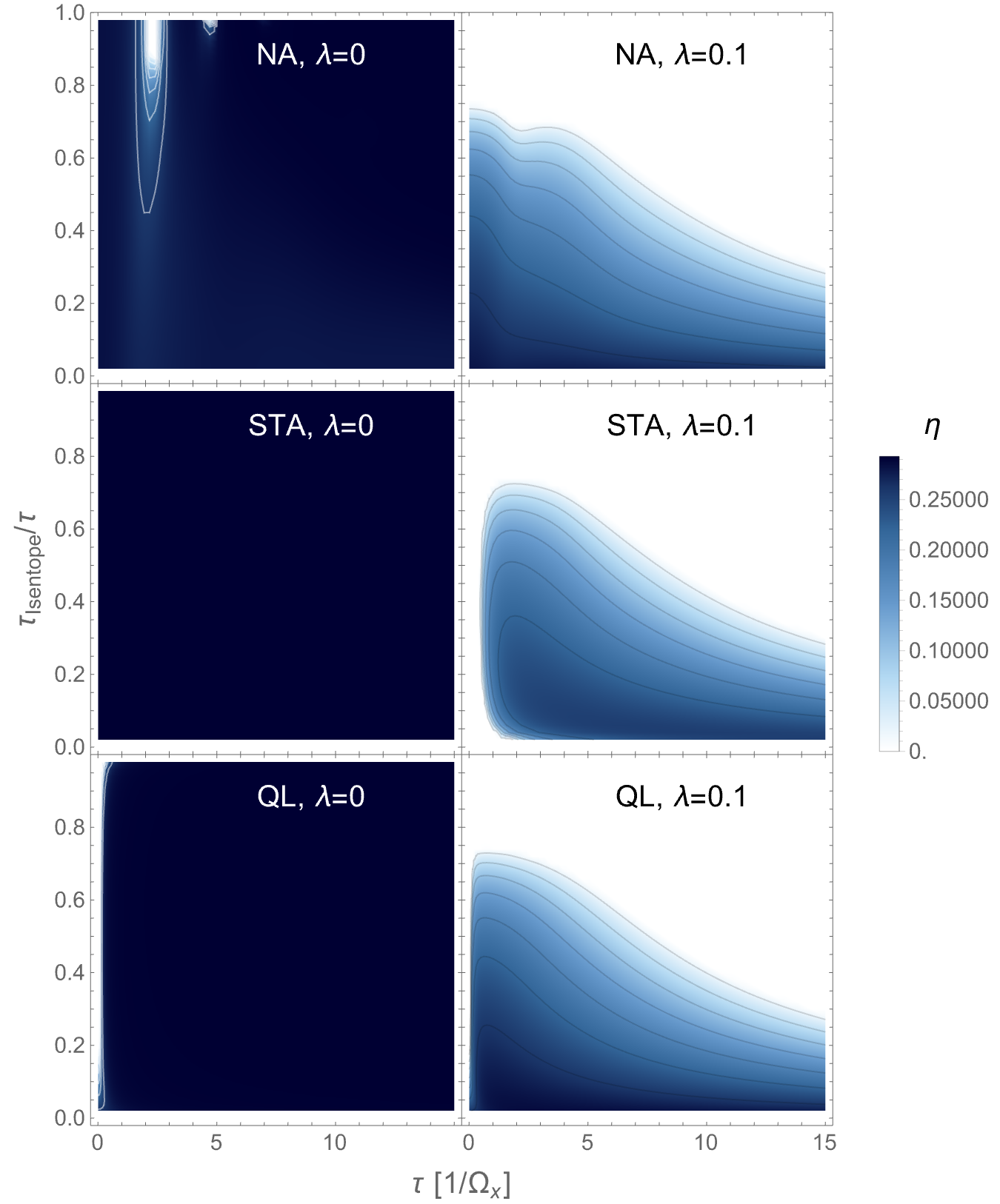}
    \end{minipage}
    
    \caption{Density plots of power [$\Omega_x^2$] (left) and efficiency (right) across total cycle times, $\tau$, (x-axis) and proportions of $\tau$ spent on the isentropic strokes (y-axis). The six setups examined have featured in Figures~\ref{fig:AvgPerformance} and \ref{fig:QL}: nonadiabatic (NA), shortcut to adiabaticity (STA) and quantum-lubricated (QL) engines, with ($\lambda=0.1$) and without ($\lambda=0$) noise on the controls. We assign the colour white to regimes where power output is negative. All other parameter values are the same as previous figures.}
    \label{fig:strokedurations}
\end{figure*}

In this section, we investigate the generalisability of the results in the previous section to different relative durations of isochores to isentropes. Sampling from a range of total cycle times, $\tau$, and taking complementary strokes to be of equal duration ($\tau_{01}=\tau_{23}$ and $\tau_{12}=\tau_{30}$), we calculate the average power and efficiency for different ratios of $\tau_\text{Isentrope}/\tau_\text{Isochore}$, where $\tau_\text{Isentrope}=\tau_{01}+\tau_{23}$ and $\tau_\text{Isochore}=\tau_{12}+\tau_{30}$, as shown in Figure~\ref{fig:strokedurations}. In agreement with Figures~\ref{fig:AvgPerformance} and \ref{fig:QL}, there is a general preference for shorter cycle times. However, when an STA is applied with noisy controls, the engine breaks down abruptly below $\tau_\text{STA}$. At even smaller $\tau$ (barely visible in Figure~\ref{fig:strokedurations}), QL techniques are likewise seen to be detrimental. The cycle time at which this occurs tends zero as $\lambda_{QL}\rightarrow\infty$. Also as expected, in Figure~\ref{fig:strokedurations} (right), the noiseless STA engine functions at Otto efficiency for all stroke duration ratios. 

Interestingly, in all setups, the average power is optimised for $\tau_\text{Isentrope} \rightarrow 0$ where isochores are much longer than isentropes. This seems to suggest, at least with the parameters chosen here, that thermalisation is more crucial to performance than transitionless driving. This effect becomes less prominent as $\tau$ increases because the system effectively reaches the thermal state at the end of each isochore regardless of the ratio of $\tau_\text{Isentrope}$ to $\tau_\text{Isochore}$. In the limit $\tau_\text{Isentrope} /\tau \rightarrow 1$, the isochores become identity maps and no heat is exchanged; thus, for all such cases in Figure~\ref{fig:strokedurations} (left), there is no power output.  

Another notable feature is the set of minima seen in NA engines for $\tau \approx 2.5$ (and again at $\tau \approx 5$), most obviously in noiseless engines when $\tau_\text{Isentrope}/\tau \rightarrow 1$. The effect is still apparent in the noisy case, though it is dwarfed by the negative impact of control noise. This feature is also non-existent for STA and QL engines (and, as will be shown in \ref{subsec:avg4PMS}, for projectively measured engines where coherence is destroyed at each vertex). Thus, we can deduce that this minima is a result of coherence.

When $\tau_{01}=\tau_{12}=\tau_{23}=\tau_{30}$, three key performance features of are visible: the existence of $\tau_\text{STA}$ (and inefficacy of QL at very small $\tau$); the degradation of noisy engines as $\tau$ increases; and the local minima seen at intermediate cycle times. For this reason, although sub-optimal for performance, we will return to the scenario where all stroke times are equal for the remainder of the paper, as it is sufficiently general to capture all interesting features.

\section{\label{sec:fluctuations} Performance Fluctuations}
\subsection{\label{subsec:4point} 4-Point Measurement Scheme}
As seen, the classical trade-off between power and efficiency is not present in an idealised quantum regime (noiseless conditions and with the use of an STA). However, there is a third measure of performance which suffers in the quantum regime, as opposed to macroscopic engines: \textit{constancy} \cite{PietzonkaSeifert}. The variance in power, a measure of constancy, can be calculated by implementing a 4-point measurement scheme (4PMS) on the vertices of the cycle to generate a power probability distribution, in a similar manner to Ref.~\cite{DenzlerLutz}.

The probability distribution of the power, $\mathcal{P}(P)$, can be derived from the probabilities of energy changes around the cycle,
\begin{align}\label{eqn:energyProbs}
     \mathcal{P}(W_{01},Q_{H},W_{23}) & = \mathcal{P}(W_{23}|Q_{H})\mathcal{P}(Q_{H}|W_{01})\mathcal{P}(W_{01}),
\end{align}
where, as per Figure~\ref{fig:ottoCycle}: $\mathcal{P}(W_{01})$ is the probability of measuring work $W_{01}$ on the compression stroke of the cycle; $\mathcal{P}(Q_{H}|W_{01})$ represents the conditional probability of measuring heat $Q_{H}$ given that, on the previous stroke, a work of $W_{01}$ was measured; and $\mathcal{P}(W_{23}|Q_{H})$ is defined similarly. If state $m$ is measured at the end of the previous stroke, $\mathcal{P}(Q_{H}|W_{01}) = \sum_s  \delta [ Q_{H} - (\Omega_s-\Omega_m)] p_{m \rightarrow s}^{\tau_{12}}$. Here, $p_{m \rightarrow s}^{\tau_{12}}$ denotes the transition probability from energy eigenstates $\ket{m}$ to $\ket{s}$ over the time $\tau_{12}$ corresponding to the hot isochore, and $\Omega_m$ denotes the eigenvalue of $\ket{m}$ which, according to Equation~\ref{eqn:ham2}, is $0$ if $\ket{m}$ is the ground state or $\Omega(t)$ if $\ket{m}$ is excited.

Stitching these conditional probabilities together we can expand Equation~\ref{eqn:energyProbs} to obtain
\begin{align}\label{eqn:energyProbs2}
     \mathcal{P}(W_{01},Q_{H},W_{23}) & = \sum_{n,m,s,v} \delta\left[W_{23} - (\Omega_s-\Omega_v) \right]\ \nonumber\\
     & \quad\quad\quad\quad \delta\left[Q_{H} - (\Omega_s-\Omega_m) \right] \nonumber\\
     & \quad\quad\quad\quad \delta\left[W_{01} - (\Omega_n-\Omega_m) \right] \nonumber\\ 
    & \quad\quad\quad\quad p_n^0 p_{n\rightarrow m}^{\tau_{01}} p_{m\rightarrow s}^{\tau_{12}} p_{s\rightarrow v}^{\tau_{23}}.
\end{align}
The above equation sums over all initial states, $n$, and their possible trajectories across the remaining 3 measurements around the cycle. The initial measurement probabilities $p_n^0$ are taken directly from the populations of the density matrix in the limit cycle at the beginning of the compression isentrope. 

We can select the probability distribution of the power by integrating over all trajectories around the cycle and imposing a definition of the power using a Dirac delta function,
\begin{align}\label{eqn:powerdistribution}
     \mathcal{P}(P) & = \int_{-\infty}^{\infty} \mathcal{P}(W_{01},Q_{12},W_{23}) \delta\left[P - \frac{W_{01}+W_{23}}{\tau}\right]\nonumber\\ 
    & \quad\quad\quad\quad dW_{01} dQ_{12} dW_{23}\nonumber\\
    & = \sum_{n,m,s,v} \delta\left[P - \frac{\Omega_n-\Omega_m+\Omega_s-\Omega_v}{\tau}\right] \nonumber\\ 
    & \quad\quad\quad\quad p_n^0 p_{n\rightarrow m}^{\tau_{01}} p_{m\rightarrow s}^{\tau_{12}} p_{s\rightarrow v}^{\tau_{23}}.
\end{align}
The moments of the power distribution can then be found by integrating over all measurement trajectories,
\begin{equation}
\label{eqn:powervariance}
    \begin{aligned}
    \langle P^\alpha \rangle = \int_{-\infty}^{\infty}\mathcal{P}(P) P^\alpha dP,
    \end{aligned}
\end{equation}
and used to define fluctuation measures like the variance.

The transition probabilities, $p_{i\rightarrow j}^{\tau_k}$, have the general form $|\bra{j(\tau_k)}V(\tau_k,0)\ket{i(0)}|^2$ where $V(t,0)$ represents the time evolution operator along the stroke. Equivalently, in Fock space, stroke propagators describe the evolution of the system over individual strokes such that $\mathcal{V}(\tau_k,0)\ket{\rho_0}\rangle = \ket{\rho_{\tau_k}}\rangle$, where $\mathcal{V}(\tau_k,0)$ is calculated from the associated Liouvillian superoperator for that stroke; $\Dot{\mathcal{V}}(t,0)=\mathcal{L}(t)\mathcal{V}(t,0)$ with $\ket{\Dot{\rho_t}}\rangle=\mathcal{L}(t)\ket{\rho_t}\rangle$. Thus, $p_{i\rightarrow j}^{\tau_k}$ can be more usefully represented using stroke propagators in Fock space,
\begin{align}\label{eqn:transitionprobabilities}
    p_{i\rightarrow j}^{\tau_k} & = \bra{j(\tau_k)}V(\tau_k,0)\ketbra{i(0)}{i(0)}V^\dagger (\tau_k,0)\ket{j(\tau_k)}\nonumber\\
    & = \bra{j(\tau_k)}V(\tau_k,0)\rho_{0}^{i} V^\dagger (\tau_k,0)\ket{j(\tau_k)}\nonumber\\
    & = \left( \bra{j(\tau_k)} \otimes \ket{j(\tau_k)}^T \right)\mathcal{V}(\tau_k,0)\ket{\rho_{0}^{i}}\rangle .
\end{align}
Taking the state at the beginning of the stroke to be collapsed, $\ketbra{i(0)}{i(0)}$, and then operating on this state with the stroke propagator over time $\tau_k$, the transition probability to the state of interest $\ket{j(\tau_k)}$ is selected using $\bra{j(\tau_k)}...\ket{j(\tau_k)}$.

When noise is present on the control, the $x^\text{th}$ moment of power (also applicable to other thermodynamic observables) is given by:
    \begin{align}\label{eqn:noisymoments}
    \mathds{E} [ \langle P^x \rangle ] & = \mathds{E} \left[ \int_{-\infty}^{\infty}  \mathcal{P}(P) P^x dP \right] \nonumber\\
    & = \sum_{n,m,s,v} \left[\frac{\Omega_n-\Omega_m+\Omega_s-\Omega_v}{\tau}\right]^x \nonumber\\ 
    & \quad\quad \mathds{E} \left[ p_n^0 \right] \mathds{E} \left[p_{n\rightarrow m}^{\tau_1} \right] \mathds{E} \left[ p_{m\rightarrow s}^{\tau_2} \right] \mathds{E} \left[ p_{s\rightarrow v}^{\tau_3} \right] .
    \end{align}
In reaching the second equality, we have made two assumptions. First, we note that the energy measurements themselves are independent of the applied control fields and are taken to be noise-free. Therefore, measurement returns energy eigenvalues, $\Omega_i$, in the (noise-averaged) basis in which we are performing our measurements. This is justified by a fundamental postulate of quantum mechanics: that the measurement of an observable by an operator, $\hat{A}$, will always return an eigenvalue $a$ of that operator, where $\hat{A} \ket{\psi} = a \ket{\psi} $. Second, as a continuation of the Markov property of white noise, where noise is uncorrelated with itself at different times, the noise-averaging acts independently on the transition probabilities. Thus, Equation~\ref{eqn:transitionprobabilities} accommodates noisy dynamics by the replacement  $\mathcal{V}(\tau_k,0) \rightarrow \mathds{E}\left[ \mathcal{V}(\tau_k,0) \right]$.


\begin{figure}
    \centering
    \includegraphics[width=0.47\textwidth]{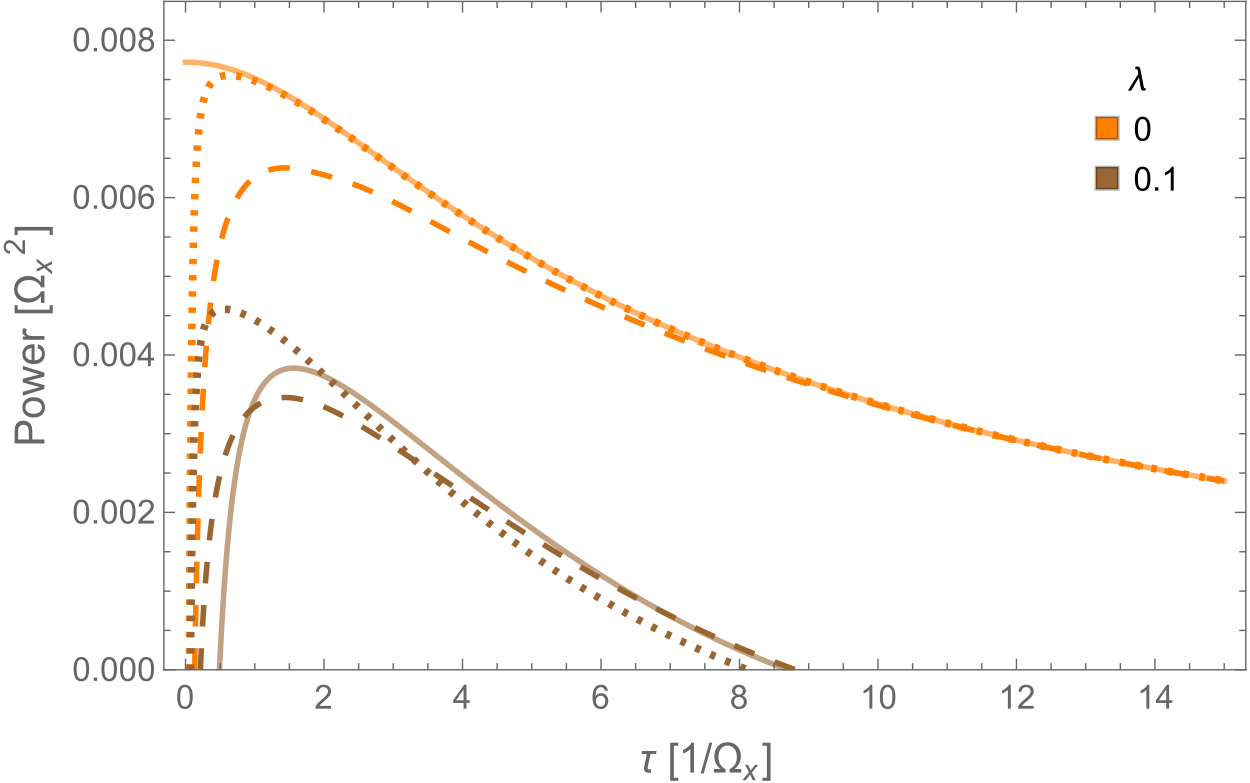}
\caption{\label{fig:AvgPerf4PMS} The power of the engine calculated using a 4PMS, in the presence of two different levels of control noise, $\lambda$, across different cycle times, $\tau$, given in units of intrinsic system energy scales, $\Omega_x$. Solid lines represent an STA engine, dotted lines a QL engine (with $\lambda_{QL}=100$), and dashed lines an NA engine. All other numeric parameters are the same as previous figures.}
\end{figure}

\subsection{Average Performance with 4PMS}\label{subsec:avg4PMS}

The average performance calculated using the 4PMS may be expected to generate significantly different results from that given in Figure~\ref{fig:AvgPerformance}. This is because coherence is destroyed as the state is collapsed by projective measurements on each vertex of the 4PMS. On the other hand, when using the energy expectation at the vertices, no such disruption occurs. Indeed, in Figure~\ref{fig:AvgPerf4PMS} we find that results differ between the approaches, most significantly for the NA engine ran quickly ($\tau < 1$). This follows from the fact that under these conditions coherence generation is largest (enhancement and slower driving reduce coherence). Interestingly, the NA engine's performance degrades when calculated using the 4PMS which suggests that, at short operation times, coherence generation can contribute positively to performance. Figure~\ref{fig:AvgPerf4PMS} also shows that QL is potentially the most effective enhancement approach when the cycle is measured according to the 4PMS. Lastly, as $\tau$ increases, results for average performance found using energy expectation values and using the 4PMS coalesce (not plotted). 

These features are in agreement with literature \cite{kosloffCoherence}, whereby at short cycle times and at certain system-bath coupling strengths, it is suggested that a quantum heat engine outperforms a classical stochastic model of a heat engine (without coherence), with the performances of the two engines becoming equal at larger cycle times when thermalisation with the bath is reached and the system fully decoheres.

The Otto engine has recently been realised experimentally in \cite{ExperimentalQHE} and \cite{PhysRevA.108.042614}, utilising projective measurements and nuclear magnetic resonance techniques on C-13 nuclei. The generated results agree qualitatively with the results of this paper. Specifically, positive work (power) output is observed only at finite cycle times ($\tau > \tau_\text{min}$). Such features could potentially originate, according to this theoretical model, due to the destruction of coherence from projective measurements (as seen in Figure~\ref{fig:AvgPerf4PMS}).

\subsection{\label{subsec:TUR}Thermodynamic Uncertainty Relation}

\begin{figure}
    \centering
    \includegraphics[width=0.47\textwidth]{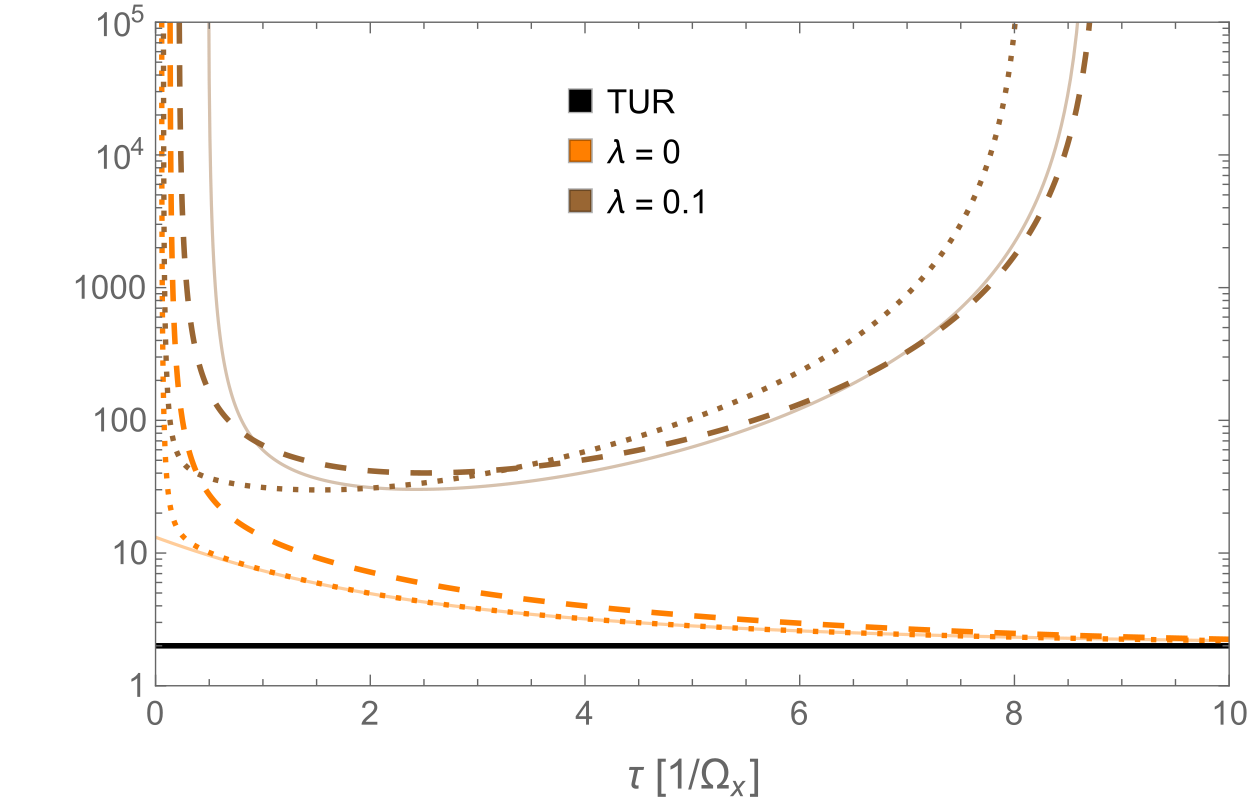}
\caption{\label{fig:TUR} The power fluctuations, $\frac{\Delta_P^2}{P_\tau^2} \Dot{ \Sigma}$, and TUR bound (equal to 2 in Equation~\ref{eqn:TUR}), at different noise levels for a single cycle of the engine across different cycle times, $\tau$, given in units of intrinsic system energy scales, $\Omega_x$. Coloured solid lines represent fluctuations of an STA engine, the dotted lines of a QL engine (with $\lambda_{QL}=100$), and dashed lines of an NA engine. All other numeric parameters are the same as previous figures.}.
\end{figure}

\begin{figure}
    \centering
    \includegraphics[width=0.47\textwidth]{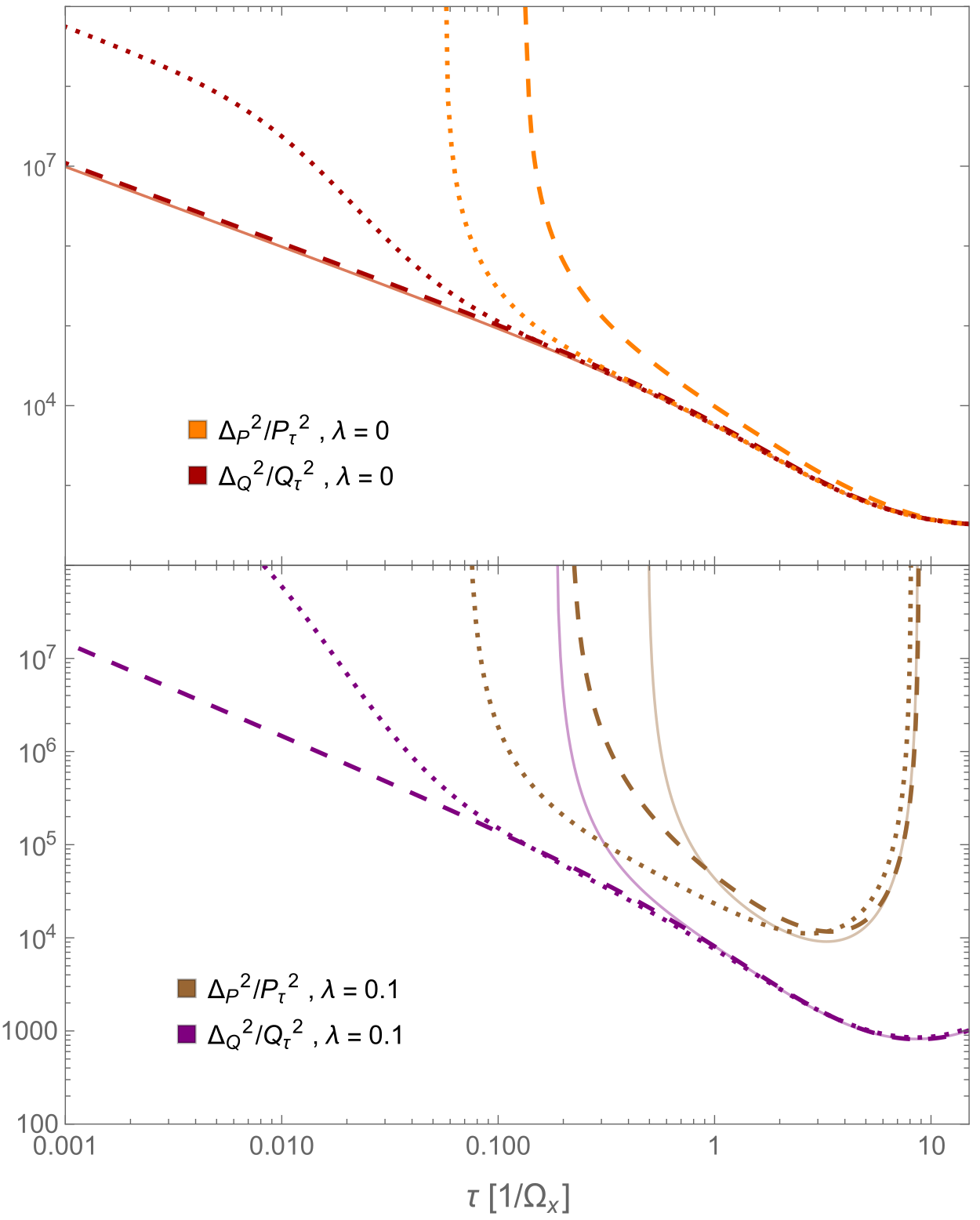}
\caption{\label{fig:QPfluctuations} The power and heat Fano factors under noiseless (top) and noisy (bottom) conditions, for a single cycle of the engine across different cycle times, $\tau$, given in units of intrinsic system energy scales, $\Omega_x$. Solid lines represent fluctuations of an STA engine, the dotted lines of a QL engine (with $\lambda_{QL}=100$), and dashed lines of an NA engine. All other parameter values are the same as previous figures.}
\end{figure}

Thermodynamic uncertainty relations (TURs) are inequalities that relate fluctuations of currents to the average entropy production, becoming increasingly relevant when driving further from equilibrium and when smaller systems are considered \cite{horowitz2020thermodynamic, LandiEntropyProduction}. With their conception in the field of statistical thermodynamics, they also apply to and have found underpinning in the quantum regime \cite{reiche2022quantum}. Here, a bound on power variance originating from a TUR is used as a consistency check for the variance calculations using the 4PMS in Equation~\ref{eqn:powervariance}, and also to observe the tightness of the bound for different scenarios. 

A relevant TUR relates the noise- and time-averaged variance in power (a work current), ${\Delta}_P^2/\tau = ( \mathds{E}\left[\langle P^2\rangle\right]-\mathds{E}\left[\langle P \rangle\right]^2)$, to the noise-averaged power, $P_\tau = \mathds{E}[\langle P\rangle]$, and the entropy production rate, $\Dot{\Sigma}$, over a cycle:
\begin{equation}
\label{eqn:TUR}
   \frac{\Delta_P^2}{P_\tau^2} \geq \frac{2}{\Dot{ \Sigma}}.
\end{equation}

This TUR is derived in accordance with Refs.~\cite{PhysRevE.103.L060103} and \cite{PhysRevLett.123.090604}, valid for time-symmetric driving and unital dynamics \cite{PhysRevE.92.032129} on the isentropic strokes, as is the case here since the opposite ramping protocol is applied on complimentary isentropes. It is worth noting that Equation~\ref{eqn:TUR} is also consistent with a bound derived for more general periodic quantum heat engines in the slow-driving limit \cite{PhysRevLett.126.210603}, where the attained corrections for time-asymmetric driving and generation of quantum friction vanish for the model under review. Furthermore, Equation~\ref{eqn:TUR} coincides with the form of TUR applicable to steady-state heat engines \cite{PietzonkaSeifert}.

The form of Equation~\ref{eqn:TUR} indicates the usefulness of the Fano factor \cite{fano1947ionization} as a measure of fluctuations, defined here as $F(P) = \Delta_P^2/P_\tau^2$. We see that smaller fluctuations are permitted by increasing the dissipation $\Dot{\Sigma}$; for instance, if we desire to reduce $\Delta_P^2$, then we must decrease the average power or increase entropy production. For a cyclic engine, $\Dot{\Sigma}$ is given by the sum of entropy fluxes from contact with heat baths and the change in von Neumann entropy, $\Delta S_{VN}$, per cycle:
\begin{equation}
\label{eqn:EntProd}
\begin{aligned}
    \Dot{\Sigma} & =  \frac{1}{\tau}\Delta S_{VN}-\underset{i \in H,C}{\sum} \beta_i \Dot{Q_i},
\end{aligned}
\end{equation}
where $\Dot{Q_i}$ is the heat current into bath $i$. For an engine operating in its limit cycle, $\Delta S_{VN} = 0$, and $\Dot{\Sigma}$ can be expressed in terms of the power, inverse temperature of the cold bath, efficiency, and Carnot efficiency of the cycle \cite{LandiEntropyProduction}. With this, we can reformulate the above bound (Equation~\ref{eqn:TUR}) in terms of efficiency, power and power variance in order to explicitly show the existence of a trade-off between the three quantities:
\begin{equation}
    \label{eqn:TUR2}
     \frac{\Delta_P^2}{P_\tau^2} \geq \frac{2 \eta}{\beta_C P_\tau (\eta_C - \eta)},
\end{equation}
where all quantities have been defined previously. 

The existence of a trade-off between average performance and constancy is consistent with Figures~\ref{fig:TUR} and \ref{fig:QPfluctuations}, where the measured fluctuations increase at cycle times when average performance is optimal ($\tau \leq 1.5$ from Figure~\ref{fig:AvgPerf4PMS}). Furthermore, in figure~\ref{fig:TUR} the noiseless and noise-averaged variances are shown to always be greater than the TUR bound as required. The measured power fluctuations for noisy controls (brown) increase sharply at $\tau \approx 8$ corresponding to the point where the engine ceases to produce a power output (see Figure~\ref{fig:AvgPerformance}). At $\tau \gtrapprox 8$, the cycle functions as an inverted heat pump, consuming work and accelerating the transport of heat from the hot to the cold bath \cite{PhysRevB.101.054513}. The performance fluctuations are only displayed for operation times when the cycle is operating successfully as a heat engine.

The variances of STA and NA engines become equal at larger times ($\tau \approx 10$), illustrated by the coalescence of dashed and solid lines of the same colour in Figures~\ref{fig:TUR} and \ref{fig:QPfluctuations}. This is in agreement with investigations \cite{funo2017universal} into the fluctuations of STA protocols without control noise. It was predicted that, although greater fluctuations occur at intermediate times during STA control stokes, at the end of the STA stroke, as calculated here, the variance should be the same as a truly adiabatic stroke without a shortcut. Conversely, most apparent in Figure~\ref{fig:TUR}, the power fluctuations of the QL engine differ from the NA engine for most cycle times. This follows from the deviating average performance between the QL and NA engines (Figure~\ref{fig:AvgPerf4PMS}). 

Another feature to note is that the TUR bound is saturated in the noiseless set-up faster than when control noise is present (this occurs at $\tau \gg 15$, not shown); in other words, the TUR bound for noisy engines is not as tight. This indicates, rather straightforwardly, that there are unnecessary fluctuations when noise exists on the controls. Lastly, for all engines, the calculated fluctuations in performance are significant with respect to the average performance, with $F(P)$ consistently greater than $1$. 

In Figure~\ref{fig:QPfluctuations}, the Fano factors of $Q_H$ and $P$ are displayed across different cycle times. As predicted in Refs.~\cite{PhysRevE.103.L060103}, the relative fluctuation of power is shown to always be greater than or equal to the relative fluctuation of input heat, $F(P)\geq F(Q_H)$, with the equality reached in the adiabatic limit. In addition, Figure~\ref{fig:QPfluctuations} highlights the short-time fluctuations of output power (represented by the orange and brown lines): with noiseless controls, an STA gives rise to the lowest fluctuations; and, with noisy controls, QL is the most reliable approach, whilst an STA increases the fluctuations beyond that which is achieved non-adiabatically.

\section{\label{sec:conc} Conclusion}
A model of a quantum Otto engine can overcome a classical trade-off between power output and working efficiency with respect to cycle operation time, when controls are noise-free. However, fluctuations in performance scale with improved average performance, consistent with bounds from the TUR. The destruction of coherence within the engine cycle via projective measurements was seen to worsen performance, especially in NA engines, aiding the conception that coherence can benefit work extraction.

An STA was shown to improve average engine performance at intermediate cycle times but was found to act detrimentally at short cycles times, due to the scaling of noise in the auxiliary field with rate of driving $\Dot{\Omega}_z(t)$. Slower engines become quasi-adiabatic and therefore the presence of an STA ceases to be beneficial. QL recovers adiabatic dynamics in the absence of control noise but fails to commute with the fluctuating system Hamiltonian when control noise is introduced and thus contributes non-trivial energetics dependent on the strength of the lubricating noise. Thus, whilst engine performance is consistently improved by quantum enhancement techniques if auxiliary controls are noiseless, this is not generally the case for noisy controls.

Further research into the generalisability of these results to other models of noise, working systems, enhancement techniques (such as noise resistant STAs \cite{levy2018noise}) and quantum devices would be of interest. Additionally, an in-depth investigation into different conceptions of heat and work would be fruitful for future studies on the thermodynamics of noisy quantum control.

\acknowledgements
This work was supported by the Engineering and Physical Sciences Research Council [grant number EP/W524347/1]. We thank Harry Miller and Obinna Abah for useful discussions.


\appendix


\section{\label{sec:appendixExp} Derivation of $\mathcal{D}_\xi$}


This appendix concerns itself with deriving the Lindblad dissipator, $\mathcal{D}_\xi$, due to noisy controls in the general time-dependent case, as in Equation~\ref{eqn:noisyDissipator}. A reminder that we are considering GWN with correlations described by
\begin{align}\label{eqn:GWNcorrelation}
  \mathcal{X}_{i j}(t-s) = \delta_{ij}\delta(t-s)  
\end{align}
with variance $\lambda$. The result for $\mathcal{D}_\xi$ is unusual in the sense that the noise strength is time-dependent, scaling with the amplitudes of the driving fields. Thus, we will focus on a closed system with time-dependent driving, an isentrope. (Note that, certainly on isochores where the driving is static, $\mathcal{D}_\xi$ behaves additively with a bath dissipator $D_B$ if the system is also coupled to a thermal reservoir. This result is not proven here but can be achieved by moving into the interaction picture with respect to the noisy dynamics and performing the below treatment of noise-averaging prior to dealing with bath terms).

If $M(\xi)$ is a functional of the real-valued noise, $\xi$, with $\mathds{E}[\xi_i(t)\xi_j(s)]=\lambda \mathcal{X}_{i j}(t-s)$, then Novikov's theorem can be written as \cite{novikov1965functionals}:
\begin{align}\label{eqn:novikov}
     \mathds{E}\left[ \xi_\alpha(t) M(\xi )\right] & = \mathds{E}\left[ \xi_\alpha (t)\right] \mathds{E}\left[ M(\xi )\right]\nonumber\\
     & + \int_0^t dt_1 \mathcal{X}_{\alpha \beta}(t-t_1) \mathds{E}\left[ \frac{\delta M(\xi)}{\delta \xi_\beta(t_1)} \right]
\end{align}
where $\frac{\delta M(\xi)}{\delta \xi_\beta(t_1)}$ represents the functional derivative of $M(\xi)$ with respect to the noise source $\xi_\beta$ at time $t_1$. We adopt from here forward the convention of implicit summation over repeated indices. 

The trajectory-dependent evolution of the state across a noisy isentrope is given by $\rho(t)$, with dynamics
\begin{equation}
    \label{eqn:rhoTraj}
    \frac{d}{dt}\rho(t) = -i [H_S(t) + H_\xi(t),\rho(t)  ]
\end{equation}
where, as in the main script, we have set $\hbar = 1$. For the time-being we use a condensed notation, defining a new Hermitian operator $K_\alpha(t)=\frac{1}{2}\Omega_\alpha(t)\sigma_\alpha$ such that $H_\xi(t)=\sum_\alpha K_\alpha \xi_\alpha(t)$. Note that any additional Hamiltonian terms representing enhancement techniques can also be divided into their deterministic and stochastic parts, retaining the form of \ref{eqn:rhoTraj}. Therefore, although not explicitly treated, results in this section apply equally to enhanced strokes.

Following \cite{PhysRevA.63.012106}, we recognise that $\rho(t)$ is a functional of $\xi_\alpha$. Utilising this fact, we can apply \ref{eqn:novikov} as
\begin{align}\label{eqn:rhoTrajNovikov}
    \mathds{E}\left[H_\xi(t) \rho (t)\right] & = \mathds{E}\left[\xi_\alpha(t) K_\alpha(t) \rho (t)\right] \nonumber\\
    & = \mathds{E}\left[ \xi_\alpha (t)\right] \mathds{E}\left[ K_\alpha(t) \rho (t)\right]\nonumber\\
    & \quad + \int_0^t dt_1 \mathcal{X}_{\alpha \beta}(t-t_1) \mathds{E}\left[ \frac{\delta K_\alpha(t) \rho (t)}{\delta \xi_\beta(t_1)} \right] \nonumber\\
     & = \int_0^t dt_1 \mathcal{X}_{\alpha \beta}(t-t_1) K_\alpha(t) \mathds{E}\left[ \frac{\delta  \rho (t)}{\delta \xi_\beta(t_1)} \right].
\end{align}
Given the Gaussian nature of $\xi_\alpha$ with zero-mean, the first term from Equation~\ref{eqn:novikov} goes to zero here. Further, since $K_\alpha (t)$ is independent of any $\xi_\alpha$, it can be taken outside the functional derivative. Similarly, 
\begin{align}\label{eqn:rhoTrajNovikovB}
    \mathds{E}\left[  \rho (t) H_\xi(t)\right] = \int_0^t dt_1 \mathcal{X}_{\alpha \beta}(t-t_1) \mathds{E}\left[ \frac{\delta  \rho (t)}{\delta \xi_\beta(t_1)} \right] K_\alpha(t) .
\end{align}

From here we follow the approach laid out in \cite{noiseFactorisation}. Noise-averaging Equation~\ref{eqn:rhoTraj} and substituting in Equations \ref{eqn:rhoTrajNovikov} and \ref{eqn:rhoTrajNovikovB},
\begin{align}\label{eqn:rhoAVG}
     \frac{d}{dt}\mathds{E}[\rho(t)] = & -i [ H_S(t),\mathds{E}[\rho(t)] ]\nonumber\\
    & -i \int_0^t dt_1 \mathcal{X}_{\alpha \beta}(t-t_1) \left[ K_\alpha(t),\mathds{E}\left[ \frac{\delta  \rho (t)}{\delta \xi_\beta(t_1)} \right]\right] .
\end{align}

We now focus our efforts on gaining an expression for $\frac{\delta  \rho (t)}{\delta \xi_\beta(t_1)}$, as all other quantities in \ref{eqn:rhoAVG} are already known. Starting from the exact solution to \ref{eqn:rhoTraj}, 
\begin{equation}
\label{eqn:rhoSolution}
\begin{aligned}
    \rho(t) = \rho(0) -i \int_0^t ds \left[ H_S(s) + H_\xi (s), \rho(s) \right],
\end{aligned}
\end{equation}
we take the functional derivative with respect to $\xi_\beta(t_1)$:
\begin{align}
    \frac{\delta  \rho (t)}{\delta \xi_\beta(t_1)}  =&  \frac{\delta  \rho (0)}{\delta \xi_\beta(t_1)}\nonumber\\
    & -i \int_0^t ds \left[ \frac{ \delta (H_S(s) + H_\xi (s))}{\delta \xi_\beta(t_1)}, \rho(s) \right]\nonumber\\
    & -i \int_0^t ds \left[ H_S(s) + H_\xi (s),\frac{\delta \rho(s)}{\delta \xi_\beta(t_1)} \right] \label{eqn:funcDeriv1}\\
     = & -i \left[ K_\beta(t_1), \rho(t_1) \right]\nonumber\\
    & -i \int_{t_1}^t ds \left[ H_S(s) + H_\xi (s),\frac{\delta \rho(s)}{\delta \xi_\beta(t_1)} \right] \label{eqn:funcDeriv2}
\end{align}
We have taken a few actions in reaching \ref{eqn:funcDeriv2}: the first term on the right-hand side of \ref{eqn:funcDeriv1} is taken to be zero as we assume that the state and the noise are initially uncorrelated; the second term is only non-zero when $\alpha=\beta$ and the integrated time $s = t_1$, and reduces to the first term of \ref{eqn:funcDeriv2}; the noise is causal on the state and not vice-versa so the state $\rho(t<t_1)$ is not affected by any noise  $\xi_\beta(t_1)$.

If we differentiate \ref{eqn:funcDeriv2} with respect to time, $t$, we get
\begin{equation}
\label{eqn:funcDerivDiff}
\begin{aligned}
    \frac{d}{dt}\frac{\delta  \rho (t)}{\delta \xi_\beta(t_1)}=
     -i  \left[ H_S(t) + H_\xi (t),\frac{\delta \rho(t)}{\delta \xi_\beta(t_1)} \right]
\end{aligned}
\end{equation}
which has exactly the same dynamics as $\rho(t)$ in Equation~\ref{eqn:rhoTraj}, other than the initial condition is given by $\frac{\delta  \rho (t_1)}{\delta \xi_\beta(t_1)} = -i \left[ K_\beta(t_1), \rho(t_1) \right]$. The time-ordered (denoted by $\mathcal{T}$) unitary evolution operator for $\rho(t)$, $U(t,t_1)=\mathcal{T} \exp \left( -i \int_{t_1}^t ds ~H_S(s) + H_\xi (s) \right)$, can then be used to generate a solution for $\frac{\delta  \rho (t)}{\delta \xi_\beta(t_1)}$ from time $t_1$ onwards:
\begin{align}\label{eqn:funcDerivEvol}
    \frac{\delta  \rho (t)}{\delta \xi_\beta(t_1)}=
     -i  U(t,t_1) [K_\beta(t_1),\rho(t_1)] U^\dagger (t,t_1).
\end{align}

Returning to the evolution of $\mathds{E}[\rho(t)]$ given by Equation~\ref{eqn:rhoAVG}, we can substitute in the noise-averaged version of \ref{eqn:funcDerivEvol} such that
\begin{align}\label{eqn:rhoAVGnonlocal}
     \frac{d}{dt}\mathds{E}[\rho(t)] = & -i [ H_S(t),\mathds{E}[\rho(t)] - \int_0^t dt_1 \mathcal{X}_{\alpha \beta}(t-t_1) \nonumber\\
     & \cross \left[ K_\alpha(t),\mathds{E}\left[   U(t,t_1) [K_\beta(t_1),\rho(t_1)] U^\dagger (t,t_1) \right]\right].
\end{align}

Ref. \cite{noiseFactorisation} goes on to resolve the time non-local nature of the above equation and perform perturbative expansions in $H_\xi (t)$. For our case, since we are only considering white noise, we can now enforce the definition of $\mathcal{X}_{\alpha \beta}(t-t_1)=$ from \ref{eqn:GWNcorrelation}. The above equation then reduces to 
\begin{equation}
\label{eqn:rhoAVGlocal}
\begin{aligned}
     \frac{d}{dt}\mathds{E}[\rho(t)] = & -i [ H_S(t),\mathds{E}[\rho(t)] \\
     & - \left[ K_\alpha(t), [K_\alpha(t),\mathds{E}\left[  \rho(t)\right]] \right],
\end{aligned}
\end{equation}
making use of the sifting property of the Dirac delta and the fact that $U(t_1,t_1)=\mathds{1}$. Thus, summing over the repeated indices and re-substituting $K_\alpha(t)=\frac{1}{2}\Omega_\alpha(t)\sigma_\alpha$, we arrive at Equation~\ref{eqn:noisyDissipator}.

\section{\label{sec:appendixDef} Defining Work and Heat}

\subsection{Standard Definitions}

In order to evaluate the thermodynamic performance of the Otto engine, the quantities of heat and work (currents) are central, from which, for example, we can define efficiency. Macroscopic distinctions between the two quantities are uncontroversial: heat is the exchange of thermal energy, and work the exchange of mechanical energy. 

Looking at a quantum system weakly coupled to a thermal bath and driven with noiseless controls, 
there is also a widely accepted distinction between heat and work \cite{Alicki_1979,gemmer2009quantum}. As such, a first law can be formulated for each stroke within the noiseless Otto engine. For stroke $j$, driven from time $t_i$ to $t_f$, the total change in energy expectation is given by
    \begin{align}\label{eqn:1stLaw}
	 E_j   =&  W_j  +  Q_j  \nonumber\\
     =& \int_{t_i}^{t_f} \text{Tr}[\Dot{H}_{S,j}(t)\rho_j(t)] ~dt + \int_{t_i}^{t_f} \text{Tr}[H_{S,j}(t)\Dot{\rho}_j(t)] ~dt,
    \end{align}
where $W_j$ and $Q_j$ represent the work done and the heat transferred to the system and can be associated with their respective terms in the second line, and $\rho_j(t)$ represents the system density matrix across stroke $j$. Since the trace of a (finite-dimensional) commutator is zero due to the cyclic property of the trace and the evolution of the noiseless system across isentropes is given by the von Neumann equation, the energy expectation on the isentropes is solely due to work exchanged with the system. Conversely, since the Hamiltonian is constant across the isochores, changes in energy expectation are entirely due to $\mathcal{D}_B$ and thus attributed to heat transfer on these strokes. Thus, this distinction assigns unitary changes in energy due to the time-dependent Hamiltonian as work, and non-unitary energy changes encoded by $\mathcal{D}_B$ as heat. 

Difficulties arise when dividing noise-induced energetics into heat and work. Noise, when considering individual trajectories, introduces an extra source of time dependence into the system Hamiltonian. Thus, according to Equation~\ref{eqn:1stLaw}, its effects could be considered as work contributions,
{\allowdisplaybreaks
\begin{align}
    \label{eqn:1stLawNoisyW}
     E_j  =& W_j  +  Q_j  +  X_j \nonumber\\
     =& \int_{t_i}^{t_f} \text{Tr}[\Dot{H}_{S,j}(t)\rho_j(t)] ~dt \nonumber\\
     & + \int_{t_i}^{t_f} \text{Tr}[H_{S,j}(t)\mathcal{D}_{B}(\rho_j(t))]~ dt \nonumber\\
     & + \int_{t_i}^{t_f} \text{Tr}[\Dot{H}_{\xi,j}(t)\rho_j(t)] ~dt
\end{align}
}where, again, terms in the first line can be associated with their respective terms in the second line.

Conversely, noisy effects could also be interpreted as a heat cost since the noise-averaged dissipator, $\mathcal{D}_{\xi}$, from Equation~\ref{eqn:noisyDissipator} gives rise to non-unitary evolution. Averaging Equation~\ref{eqn:1stLaw} over noise realisations and using Equation~\ref{eqn:evolution1} for $\mathds{E}\left[\Dot{\rho}_j(t)\right]$ (or equivalently, applying Novikov's theorem \cite{novikov1965functionals} to Equation~\ref{eqn:1stLawNoisyW}), we get
{\allowdisplaybreaks
\begin{align}
    \label{eqn:1stLawNoisy}
    \mathds{E}\left[ E_j \right]  =&  \mathds{E}\left[W_j\right]  +  \mathds{E}\left[Q_j\right]  +  \mathds{E}\left[X_j\right]  \nonumber\\
     =& \int_{t_i}^{t_f} \text{Tr}[\Dot{H}_{S,j}(t)\mathds{E}\left[\rho_j(t)\right]] ~dt \nonumber\\
     & + \int_{t_i}^{t_f} \text{Tr}[H_{S,j}(t)\mathcal{D}_{B}(\mathds{E}\left[\rho_j(t)\right])]~ dt \nonumber\\
     & + \int_{t_i}^{t_f} \text{Tr}[H_{S,j}(t)\mathcal{D}_{\xi}(\mathds{E}\left[\rho_j(t)\right])]~ dt
\end{align}}
The subject of this appendix is the classification of the noise contribution $X$ into heat and work. Here, we will only consider the energetics due to white noise sources but some of the discussion and conclusions may apply to other, more involved noise processes.
\begin{figure*}
    \centering
        \includegraphics[width=\linewidth]{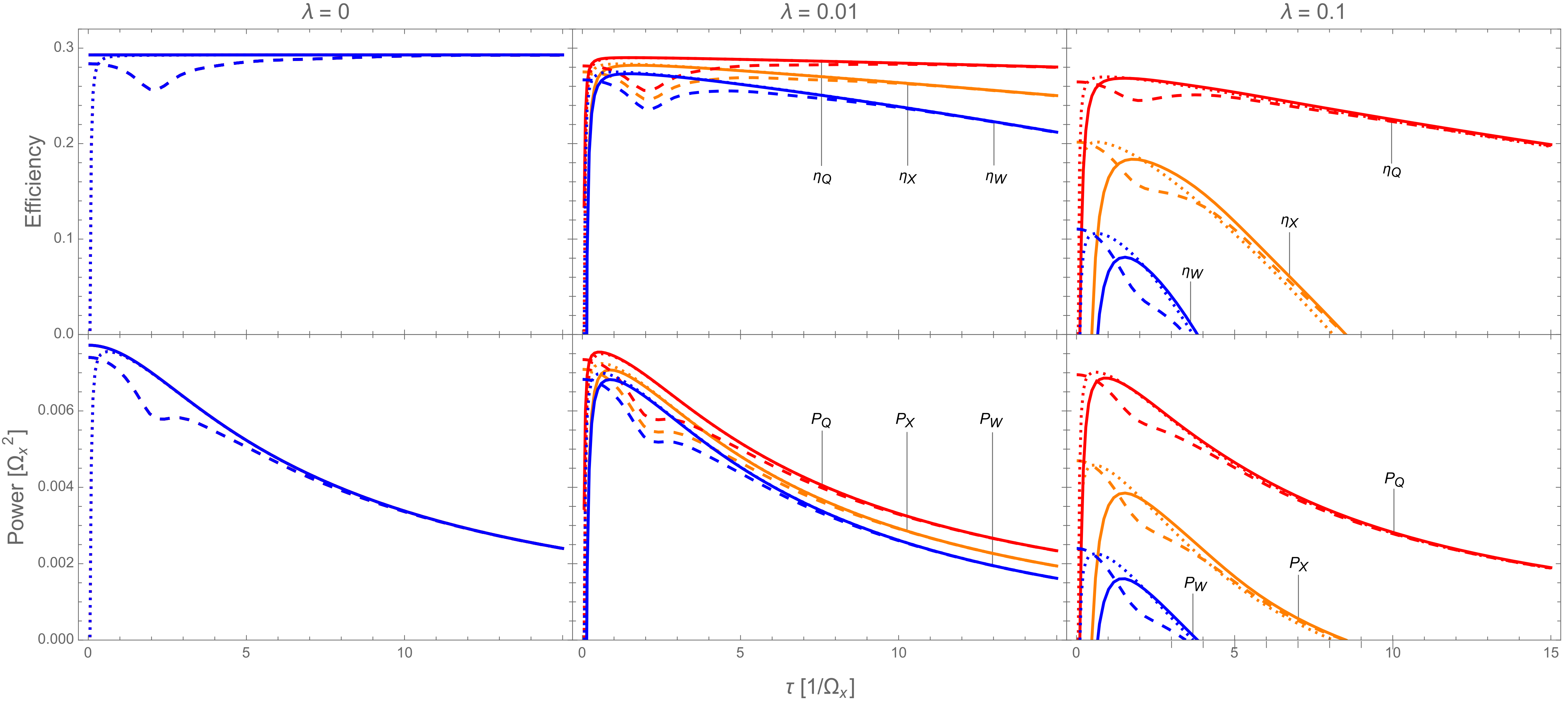}
    
    \caption{Efficiency (top row) and Power (bottom row) across total cycle times for 3 different noise levels $\lambda = (0,0.01,0.1)$ (left, middle, right column). The different colour lines refer to different classifications of the noise-induced energetics, $X$, and different dash styles refer to different enhancement techniques (STA represented by Solid lines, NA by Dashed, QL by Dotted). All numeric parameters are the same as previous figures.}
    \label{fig:XClassification}
\end{figure*}

\subsection{$X$ as Heat or Work} 
There could be (at least) three reasons for considering $X$ as a form of heat:
\begin{enumerate}
    \item[(a)~] Exemplified by its dependence on the dissipator $\mathcal{D}_{\xi}$, $X$ is a result of non-unitary, irreversible dynamics after noise-averaging - qualities typical of heat flow; \\
    \item[(b)~] $\mathds{E}\left[X_j\right]$ from Equation~(\ref{eqn:1stLawNoisy}) parallels neatly with the last term in Equation~\ref{eqn:1stLaw}, which is associated with heat in the weak coupling noiseless regime;\\
    \item[(c)~] Outside of the limit cycle of the engine, there is a change in entropy between subsequent cycles. Therefore, $X$, due to its associated entropy production, is similar to heat.\\
\end{enumerate}
Thus, there is motivation to conclude that $X$ ought to be considered as a heat cost. However, such a conclusion is not without opposition. 

Regarding (a), the non-unitary and irreversible nature of control noise, whilst a common feature of quantum heat contributions, does not necessitate it as a type of heat. In fact, \textit{irreversible} work (due to quantum friction, control noise or measurement back-action, for example) is an established concept in quantum thermodynamics \cite{plastinaFriction}, and the related field of stochastic thermodynamics \cite{PhysRevLett.78.2690,ahmadi2021irreversible}.

In response to (b), whilst it is convenient to stick simply with Equation~\ref{eqn:1stLaw}, convenience alone is insufficient reason and there seems to lack a microscopic motivation for assigning all dissipative action (including from averaging over fluctuating controls) as heat. Further, Equation~\ref{eqn:1stLaw} was not designed to be fully general. Separate from the issue of noisy energetics, there are other instances when this classification breaks down; for example, non-Markovian evolution induced by strong-coupling to a bath \cite{kato2016quantum,strasberg2016nonequilibrium,shubrook2025non,PhysRevE.101.052129,PhysRevE.95.032139}. 

In reference to (c), it is not generally true that entropy production should always be connected to heat, only that it be sourced from an irreversible process (of which control noise is one). Entropy \textit{flow}, however, is always a manifestation of heat exchange. White noise, having an infinite effective temperature, has zero entropy flow. Therefore, there is no reason to consider noise-induced energetic changes as heat exchanges. 

Looking at the issue from another perspective, $X$, at least in most setups (including the one studied in this paper), is not consumed as a resource and does not need replenishing. Meanwhile, heat sources do require recharging, thus have an associated heat \textit{cost}. Similar to how the heat exchange with the cold bath, assumed to be ambient, does not feature in the definition of efficiency as there is no cost in using it, so too can we consider noise on the controls as an ambient phenomenon and so ignore any related costs in our definition of efficiency. Overall, while there is some basic intuition for such a position, the view that $X$ is solely a heat cost appears unfavourable.

\subsection{$X$ as Work}

As an alternative to the previous approach, it may be preferable to attribute energy fluxes due to noise as a form of work. The intuition here is exemplified in the scenario of a driven system, isolated from any thermal environments. If the controls are noiseless, energy measurements on the system equate straightforwardly to the work done to or by the system as in Equation~\ref{eqn:1stLaw}. If the controls experience noise, under this approach, we still consider energy measurements as capturing the work done, but accept that this quantity is now stochastic (and likely degraded since rapid changes in controls tend to induce non-adiabatic transitions in the quantum system). As previously mentioned, this quantity is established, being referred to in the literature under different labels as fluctuating \cite{PhysRevX.6.041017,Bäumer2018}, irreversible \cite{plastinaFriction,PhysRevLett.78.2690,ahmadi2021irreversible}, dissipated \cite{PhysRevLett.123.230603,PhysRevLett.133.070405}, or stochastic \cite{sone2021jarzynski,PhysRevResearch.5.043085} work. 

Moreover, considering small time scales of individual noise trajectories, noise adds an extra source of time dependence in the system controls, consistent with Equation~(\ref{eqn:1stLaw}) before noise-averaging. This agrees with the Stratonovich interpretation of white noise which assumes that noise is a continuous function and becomes ``white'' by taking the infinitesimal limit of its self-correlation timescale \cite{moon2014interpretation}. Therefore, assigning $X$ as a work contribution has microscopic justification.

Contrary to the interpretation of $X$ as a form of work, points (a)–(c) can again be invoked to support the classification of $X$ as heat. Analogous to how energy exchanges with a cold ambient bath are excluded from efficiency definitions — despite their entropic contributions — there is also the possibility that $X$ ought not be considered a resource at all, and therefore should not be included in definitions of either heat or work. While this perspective will not be explored further here, it serves to illustrate that the classification of $X$ need not be confined to the binary of heat versus work.

It is also worth acknowledging a more practical concern with such binary classifications: their lack experimental applicability. For example, if $X$ is considered a work-related quantity, when looking at dynamics of a system experiencing noisy control and coupled to a thermal reservoir (such as the isochores in the presented model of a quantum Otto engine), there are now work contributions despite a noise-averaged Hamiltonian that is constant. This means that in order to properly quantify the work statistics of such a stroke, energy measurements of a large bosonic heat bath are required to the precision of energy scales of the working quantum system. Furthermore, for a quantum heat engine, both the hot bath and the cold ambient bath require measurement. Precise measurements of an ambient environment are potentially even more implausible than an engineered bath. Here, we take this concern seriously and proceed to evaluate energy exchange statistics using an experimentally realisable measurement procedure.

\subsection{$X$ using Projective Measurements on the System}

Quantum work statistics depend highly on the work extraction or measurement procedure. For the purposes of this paper, and due to the impracticability in precisely measuring heat reservoirs, we adopt a division of $X$ into heat and work based on the perspective of an observer who is equipped with the ability to projectively measure the system (only). The observer, assuming their controls to be almost-perfect (affected by noise only weakly), records energy changes in  the system along thermal strokes (where the system is connected to a bath and held with static controls) as heat and changes along driving strokes (where the system is isolated and driven by time-dependent controls) as work.

We recognise that there are limitations to the classification of $X$ adopted here; for example, when generalising to situations where the system is intentionally driven and in contact with a bath. Yet, even if lacking complete generality, the adopted treatment of $X$ suffices for our current purposes and, with an emphasis on experimental realisability, gives rise to testable predictions.

To fully address this matter, we have conducted calculations for the expectation values of power and efficiency (comparable to Figures (\ref{fig:AvgPerformance}) and (\ref{fig:QL}) in the main body) using the alternative categorisations of $X$ mentioned above. This was done by finding the limit cycle dynamics for the each stroke and setup and then splitting the total energy change of each stroke $j \in J =\{01,12,23,30\}$ into work, heat and noise contributions:
{\allowdisplaybreaks
\begin{align}\label{eqn:WQXj}
    W_{j}& = \int_0^{\tau_j} \Tr{\Dot{H}_{S,j}(t)\rho_j(t)} ~dt, \\
    Q_j & = \int_0^{\tau_j} \Tr{H_{S,j}(t) \mathcal{D}_B \rho_j(t)}~ dt,\\
    X_j & = \int_0^{\tau_j} \Tr{H_{S,j}(t) \mathcal{D}_\xi \rho_j(t)}~ dt. 
\end{align}}From these quantities, using the stroke labels from Section~\ref{sec:model}, we can calculate three definitions for the total heat and work for a complete cycle:
{\allowdisplaybreaks
\begin{align}\label{eqn:WQXtot}
    W_W& = - \sum_{j\in J} ( W_j + X_j ),\\
    W_Q &= - \sum_{j\in J} W_j ,\\
    W_X &= - \sum_{j\in J} W_j - (X_{01}+X_{23}),\\
    Q_W &= \sum_{j\in J} Q_j, \\
    Q_Q &= \sum_{j\in J} ( Q_j + X_j ), \\
    Q_X &= \sum_{j\in J} Q_j + (X_{12}+X_{30})
\end{align}}where the subscript $_W$ denotes quantities evaluated under the assumption that $X$ is treated as work, $_Q$ corresponds to $X$ considered as heat, and $_X$ refers to the approach adopted in accordance with projective measurements on the system.

The results presented in Figure~\ref{fig:XClassification} show the average power and efficiency \footnote{Efficiencies are generally defined as $\eta_A = W_A/Q_A$. However, even when $X$ is considered as a work cost, in Ref. \cite{kosloff2017quantum} it also appears in the denominator of the efficiency such that $\eta_K = W_W/Q_Q$. By construction, $\eta_K$ has the same domain of utility as $\eta_W$ (both are zero when $W_W$ is zero) and also has a reduced magnitude, $|\eta_K(t)| \leq |\eta_W (t)| ~ \forall t$. With no notable qualitative differences other than these, results for $\eta_K$ have not been shown in Figure~\ref{fig:XClassification}.}
for different engines (NA, STA, and QL), total cycle times ($\tau$), and different classifications of $X$ as work, heat, or defined by the 4PMS. In general, considering $X$ as a heat cost predicts greater power and efficiency and considering $X$ as work damages the performance expectations. The deviations in average performance between the approaches scale with the noise strength but with no peculiarities or notable qualitative differences. Therefore, there is motivation to believe that the results for performance and fluctuations using the 4PMS in the main body of the paper would also differ insubstantially if alternative classifications of $X$ were examined. Thus, the results in Section~\ref{sec:fluctuations} serve as indicators of the general statistical behaviour of quantum Otto engines regardless of the choice of $X$ classification. 

\bibliography{apssamp}

\end{document}